\documentclass[sort&compress,square,comma,authoryear]{aa}  

\usepackage{color}
\usepackage{graphicx}
\usepackage{graphicx}
\usepackage{dblfloatfix} 
\usepackage{adjustbox}
\usepackage{titlesec}
\usepackage[symbol]{footmisc}
\usepackage{gensymb}
\def\SPSB#1#2{\rlap{\textsuperscript{\textcolor{black}{#1}}}\SB{#2}}

\def\SB#1{\textsubscript{\textcolor{black}{#1}}}
\usepackage{txfonts}
\begin{document} 

   \title{Unveiling the traits of massive young stellar objects through a multi-scale survey\footnotemark}
   \titlerunning{A multi-scale survey of MYSOs}


   \author{A. J. Frost\inst{1}, R. D. Oudmaijer\inst{2}, W. J. de Wit \inst{3} \and S. L. Lumsden\inst{2}
          }

   \institute{Institute of Astronomy, KU Leuven, Celestijnenlaan 200D, Leuven, 3001, Belgium\\
   \email{afrost274@gmail.com}
         \and
         School of Physics and Astronomy, University of Leeds, Leeds LS2 9JT, UK
         \and
             European Southern Observatory, Alonso de Cordova 3107, Vitacura, Santiago, Chile
             }

   \date{Received 23rd October 2020 / Accepted 8th February 2021}

 
  \abstract
   {The rarity and deeply embedded nature of young massive stars has limited the understanding of the formation of stars with masses larger than 8M$_{\odot}$. Previous work has shown that complementing spectral energy distributions with interferometric and imaging data can probe the circumstellar environments of massive young stellar objects (MYSOs) well. However, complex studies of single objects often use different approaches in their analysis. Therefore the results of these studies cannot be directly compared.}
   {This work aims to obtain the physical characteristics of a sample of MYSOs at $\sim$0.01" scales, at $\sim$0.1" scales, and as a whole, which enables us to compare the characteristics of the sources.}
   {We apply the same multi-scale method and analysis to a sample of MYSOs. High-resolution interferometric data (MIDI/VLTI), near-diffraction-limited imaging data (VISIR/VLT, COMICS/Subaru), and a multi-wavelength spectral energy distribution are combined. By fitting simulated observables derived from 2.5D radiative transfer models of disk-outflow-envelope systems to our observations, the properties of the MYSOs are constrained.}
   {We find that the observables of all the MYSOs can be reproduced by models with disk-outflow-envelope geometries, analogous to the Class I geometry associated with low-mass protostars. The characteristics of the envelopes and the cavities within them are very similar across our sample. On the other hand, the disks seem to differ between the objects, in particular with regards to what we interpret as evidence of complex structures and inner holes.}
   {The MYSOs of this sample have similar large-scale geometries, but variance is observed among their disk properties. This is comparable to the morphologies observed for low-mass young stellar objects. 
    A strong correlation is found between the luminosity of the central MYSO and the size of the transition disk-like inner hole for the MYSOs, implying that photoevaporation or the presence of binary companions may be the cause.
   }


   \maketitle
%
\renewcommand{\thefootnote}{\fnsymbol{footnote}}
\footnotetext[1]{Based on observations made with ESO telescopes at the Paranal Observatory under programme IDs 097.C-0320(A), 75.C-0755, 74.C-0389, 082.C-899(A), 076.C-0725(B), 84.C-0183(D), 84.C-0183(C), 84.C-0183(B), 84.C-0183(B), 90.C-0717(A), 89.C-0968(A), 82.C-0899(A), 84.C-1072(A), 84.C-1072(B), 60.A-9224(A), 273.C-5044(A), 74.C-0389(B), 75.C-0755(A), 75.C-0755(B), 75.C-0755(A), 75.C-0755(B), 77.C-0440(A) and 381.C-0607(A).}

\section{Introduction}

Massive stars (those with masses greater than or equal to 8M$_{\odot}$) shape the Universe at all scales, with their winds, outflows, and supernovae (SNe) affecting phenomena such as interstellar chemistry, the nature of molecular clouds, and the morphology of galactic superwinds \citep{leith}. However, despite the importance of massive stars, their formation is still not completely understood. While forming low-mass stars lose their natal material (envelope) faster than their disk, revealing a clear star-disk system, young massive stars remain deeply embedded. This, combined with the fact that they are numerically rare, makes them more difficult to investigate. One crucial stage of massive star formation is the massive young stellar object (MYSO). Here, the forming star is deeply embedded in its envelope, rendering it optically invisible. Energy does escape an MYSO at micron wavelengths and longer with the most luminosity emitted at around 100$\mu$m, meaning they are IR bright. Despite the high bolometric luminosity of MYSOs, there is no ionising radiation. This implies that the star is still accreting at a high rate, making the MYSO a key stage of the massive star formation process. With current instrumentation available for the IR and (sub-)millimetre wavelengths, similar spatial resolutions ($\sim$0.01") can be reached, allowing direct comparison of morphologies of distinct but related emitting regions and their physical processes \citep{bw}. 

Disks, a key component of the low-mass star formation process, have recently been detected around MYSOs (e.g. \citealt{kraus}, \citealt{john}, \citealt{jilee18}, \citealt{maudafgl1}). \citet{car18} detected evidence of accretion around the MYSO S255 NIRS 3, observing a radio burst after a maser and IR burst and interpreting this as an accretion event of the order 10$^{-3}$M$_{\odot}$yr$^{-1}$. This implies that the detected disks and outflows observed around MYSOs are engaged in the accretion process, as they are in low-mass star formation. The low-mass star formation process has been assigned four distinct evolutionary stages based on the evolution of the envelope and disk \citep{shu87}. However, this has not been achieved for massive star formation as massive protostellar disk evolution in particular is yet to be constrained. Now that they have been confirmed to exist, determining the nature of massive protostellar disks and their role in the accretion process for massive protostars can be investigated.

Various studies have aimed to resolve the near- and mid-IR emitting regions of MYSOs by means of IR interferometry, in particular using the Very Large Telescope \& Interferometer (VLTI). For example, \citet{wit10} determined the importance of the bipolar outflow cavities as a source of mid-IR emission through the study of the MYSO W33A. Alternatively, \citet{wheel} and \citet{witcomics} used Q-band imaging to investigate the envelopes, allowing them to probe cooler regions of the protostellar environments. These studies made use of radiative transfer modelling in order to compare synthetic images to observed ones. These studies also considered spectral energy distributions (SEDs), which provide a multi-wavelength view of MYSOs but only provide, at best, indirect spatial information. As such, they must be combined with spatially resolved observations to provide geometrical information on the MYSOs. 

Another potential source of IR photons are the warm circumstellar disks and the dust at sublimation temperature. In \citealt{frost} (hereafter Paper I), we combined the
above techniques to provide a multi-scale view of the MYSO G305.20+0.21 (hereafter G305). By simultaneously fitting interferometric data between $\sim$7-13$\mu$m, Q-band direct imaging data at
$\sim$20$\mu$m, and an SED with a 2.5D radiative transfer model, the characteristics of the MYSO were obtained. We found that the broad $\sim$10$\mu$ emission traces not only the cavity emission \citep{wit10} but is also sensitive to the emission originating in the disk. This is in agreement with previous work by \citet{boley13}, who found that their data from the MID-infrared Interferometer (MIDI, VLTI) can trace both disks and outflows. Furthermore, the inner dust radius, corresponding the inner dust disk boundary, had to be over three times the sublimation radius to satisfy the observables. The SED is dominated by emission coming from the outflow cavities and envelope, constituting reprocessed emission from the star and hot inner disk regions. The fact that the disk is likely to have an inner hole and has rather low envelope density suggests that this source is more evolved in its formation and is more comparable to a transition disk phase, similar to those seen around low-mass stars.

In this paper, we apply the methodology of Paper I to an additional seven MYSOs. In total, this constitutes the largest MYSO sample to date to be subject to a multi-wavelength detailed analysis of this sort. In doing so, we obtain a detailed set of characteristics for each MYSO that can be consistently compared to each of the others, which has been lacking in the literature. In Section 2, we describe the observations. In Section 3, the methodology is briefly reiterated. In Section 4, the main results of the fitting process are described (an in-depth presentation of each source is provided in Appendix A). In Section 5, we discuss the results of the sample. Finally, we summarise and conclude in Section 6. 

    \begin{figure*}
   \centering
   \includegraphics[width=170mm]{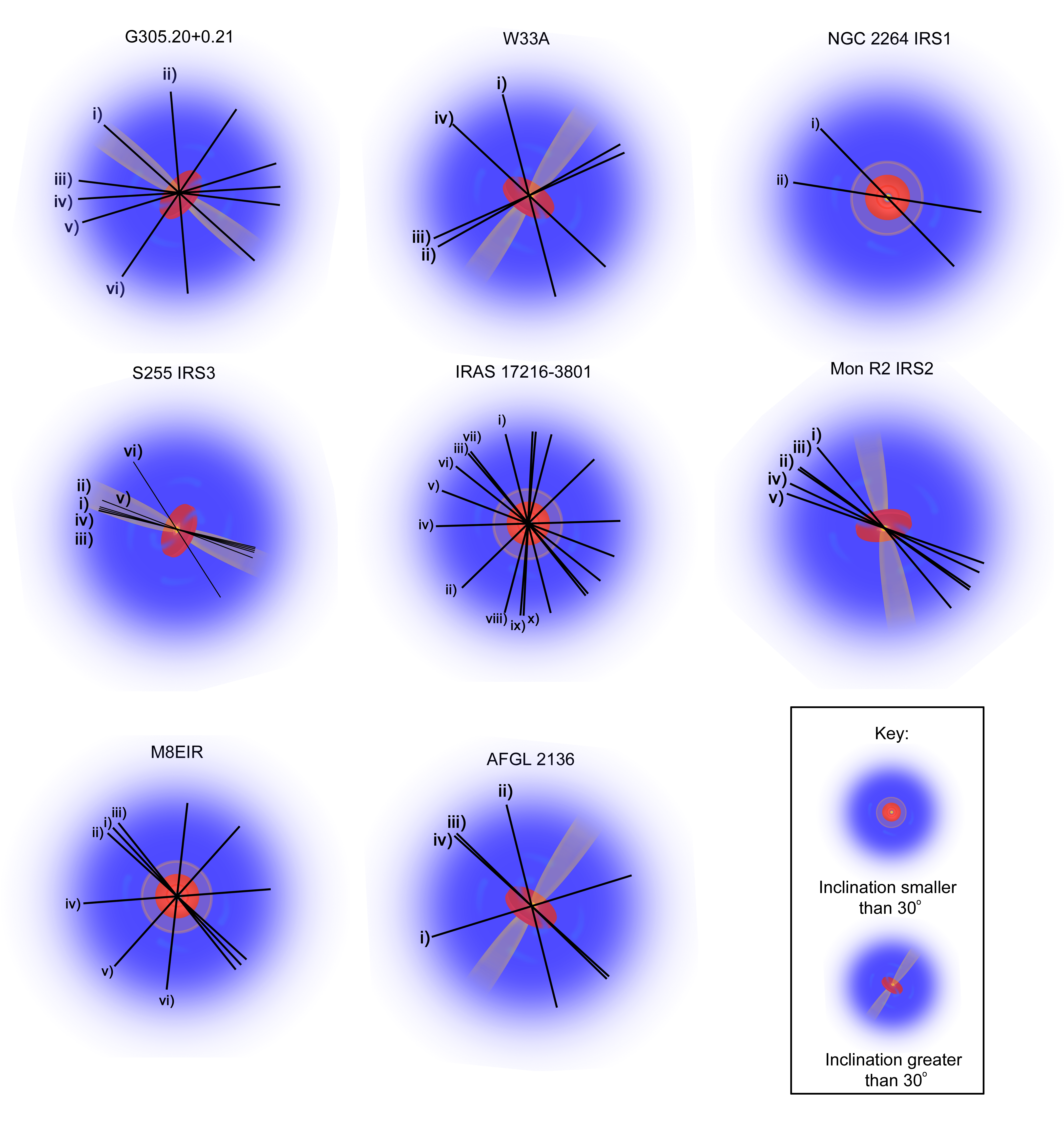}
   \caption{Schematics of the suspected geometries of each of the sources with the position angles of each configuration of their MIDI data additionally shown. G305.20+0.21, the subject of Paper I, is also shown. The envelope is shown in blue, the cavity in yellow and the disk in red. Two different schematics are shown, one that applies to sources that have inclinations of less than 30 degrees. In this case, most of the cavity emission will be along the same region of the disk in the line of sight. The other schematic represents sources inclined to the point where cavity emission extends beyond the disk in the line of sight.}
   \label{allmidiconfigs}
   \end{figure*}

\section{Observations}

The sample of MYSOs consists of eight sources located at a variety of distances, listed in Table \ref{mysosamplelist}. The sources were chosen based on the availability of extensive observational data coverage. 
The types of dataset used between this paper and Paper I are the same, although the exact instrument and source of the data varies slightly, the details of which we describe here.

\subsection{N-band interferometric data}

An important source for the mid-IR interferometric data of this sample of MYSOs is \citet{boley13}. These data were all collected with the MIDI instrument. MIDI is a two-telescope beam combiner that operates at N-band and delivers spectrally dispersed interferometric observables. It was fed by the VLT interferometer until its decommissioning in 2015. 
 For all except two sources, the MIDI data used were reduced and previously presented in \citet{boley13}. In addition to the data from \citet{boley13}, one configuration of MIDI data from \citet{grell} was used for one source (NGC 2264 IRS1) and the MIDI data from \citet{wit10} was used for W33A. The data of W33A were observed using a similar sequence to the \citet{boley13} data, using the HIGHSENS mode (where correlated and total flux measurements are taken separately and the photometric observations immediately follow the interferometric measurements to produce the final visibilities) and the same prism. They used calibrators from \citet{cohen} as well as a different version (1.6) of the data reduction software MIA+EWS, but the observing procedure followed was very similar to that of \citet{boley13}. For the \citet{grell} data an estimate of the sky background was made and subtracted over an area corresponding to the size of the mask over which the correlated spectra were extracted. 
The data from \citet{grell} were taken in the SCIPHOT mode using a grism with a spectral resolution of $\lambda$/$\Delta{}\lambda{}$ $\approx$ 230. 

Overall the MIDI data were obtained using a homogeneous set up, with the main difference being the specific u-v coverage of each set of observations. This depends on the specific observing configuration, and the areas probed for each source are illustrated in Figure \ref{allmidiconfigs}. A full list of the MIDI observations used in our work can be found in Appendix B. 

  \begin{table*}
\caption{List of the MYSOs included in this study with a column stating whether VISIR or COMICS data were used in the fitting process. \newline \textbf{References:} (1) \citet{rmslum}; (2) \citet{rmsmott10}; (3) \citep{mottlum}; (4) \citet{immer}; (5) \citet{grell}; (6) \citet{kamezaki}; (7) \citep{s255l}; (8) \citet{s255d}; (9) \citet{hen92}; (10) \citet{mondist}; (11) \citet{kraus17}; (12) \citet{boley13}; (13) \citet{linz09}; (14) \citet{damgaia}; (15) \citet{urq12}; (16) \citet{urq14}.}              
\label{mysosamplelist}      
\centering         
\begin{tabular}{c c c c c c}          
\hline\hline                        
Name & RA (J2000) & Dec. (J2000)& log($\frac{L}{L_{\odot}}$) & Distance & VISIR (V) or COMICS (C)?\\\ 
 & (h:m:s) & (d:m:s) & &(kpc) & \\
\hline                                   
G305.20+0.21 & 13:11:10.45 & -62:34:38.6 & 4.7\textsuperscript{1} & 4.0\textsuperscript{2} & V \\
W33A & 18:14:39.0 & -17:52:03 & 4.5\textsuperscript{3} & 2.4\textsuperscript{4} & C \\
NGC 2264 IRS1 & 06:41:10.15 & +09:29:33.6 & 3.6\textsuperscript{5} & 0.7\textsuperscript{6} & C \\
S255 IRS3 & 06:12:54.02	& 17:59:23.60 & 4.7\textsuperscript{7} & 1.8\textsuperscript{8} & C \\
Mon R2 IRS2 & 06:07:45.8 & -06:22:53.2 & 3.8\textsuperscript{9} & 0.8\textsuperscript{10} & C \\
IRAS 17216-3801 & 17:25:06.51 & -38:04:00.4 & 4.8\textsuperscript{11} & 3.1\textsuperscript{12} & V\\
M8EIR & 18:04:53.18 & -24:26:41.4 & 3.8\textsuperscript{13} & 1.3\textsuperscript{14} & C\\
AFGL 2136 & 18:22:26.38 & -13:30:12.0 & 5\textsuperscript{1} & 2.2\textsuperscript{15,16} & C \\
\hline                                             
\end{tabular}
\end{table*}

\subsection{Q-band imaging data}

This paper uses single-dish Q-band data ($\sim$20$\mu$n) obtained with the VISIR instrument at the VLT \citep{lag} and the COMICS instrument at the Subaru telescope. VISIR is a mid-IR imager and spectrograph (mounted on Unit Telescope 3 (UT3) of the VLT) and was used in service mode to observe the sources G305 and IRAS 17216-3801 as a part of ESO run 097.C-0320(A) (PI: A. J. Frost). The observations were taken with the Q-band filter which has a central wavelength of 19.5$\mu$m and a half-band-width of 0.4$\mu$m, tracing cooler material than that of MIDI and therefore providing an alternative view of the protostellar environments. The data reduction and further details on this VISIR data were referenced in Paper I. 
COMICS observations were used at a central wavelength of 24.$\mu$m for the remaining sources, somewhat longer than the VISIR observations but still sensitive to cavity wall and envelope emission. Images were taken with a pixel size of 0.13 $\times$ 0.13 arcsec$^2$ and a field of view of approximately 40 $\times$ 30 arcsec$^2$. 
The COMICS data presented here were reduced as part of \citet{witcomics} and processed with IRAF \citep{iraf}. 

All Q-band observations were accompanied by PSF standards and all target objects are spatially resolved. HD 123139 was observed as PSF standard on the night of the VISIR observations with a measured FWHM of 0.48" and $\alpha$ Tau was used as the PSF during the post-processing of the COMICS observations and has a FWHM of 0.34". We performed aperture photometry on the Q-band images in a standard way comparing the flux behaviour of flux standard throughout the night. In our modelling, we reduce the 2D images to azimuthal averaged radial profiles, which are then compared to synthetic data.





\subsection{Spectral energy distributions}

The fluxes for the SEDs were compiled on an object-by-object basis. 
The Red MSX Source (RMS) Survey \citep{rmslum} database was used in addition to the literature to create an observational SED for each object. 
The literature was consulted to ascertain whether, as was the case for G305, other sources could be contaminating various fluxes, and if this was the case such measurements were not considered in the fitting. All the included SED fluxes are listed in Appendix C.


\section{Methodology}

The methods of this work closely follow those of Paper I. In short, the multi-wavelength dataset described in Section 2 is used to probe the protostellar environments of MYSOs at 10mas and 100mas scales and as a whole. By fitting one RT model to all these observations simultaneously, the characteristics of the MYSO environment at these multiple scales are probed, providing a multi-scale view of the MYSOs. 

HOCHUNK-3D \citep{whitney} is the code used to generate the RT models from which the required simulated observables are extracted. The model consists of two main geometrical components; an envelope with bipolar outflow cavities and a disk, which surround a central object. Polynomial-shaped or streamlined cavities present the two main geometrical options for the bipolar outflow cavities, and two envelope density distributions - power-law and Ulrich \citep{ulrich} (a rotating and infalling envelope) - are explored. The disk is described by two separate density components. One is a large-grain disk component and the other a small grains disk component, which are assigned different dust types. The base stellar parameters sampled are the stellar radius, mass, and temperature. We find that it is the stellar luminosity, as opposed to the individual radius and temperature, which affect our observables, so this is discussed in later sections. The disk is modelled via two separate disk components, a large-grains disk (disk 1 - described by Model 1 from \citealt{wood02}) and a small grains disk (disk 2 - described by the interstellar medium (ISM) grain model of \citealt{kim}). For each component the sample parameters were the inner radius, outer radius, and scale-height. In Paper I, only the large-grain disk component was included; in this work, both disk components are used. This allows for different geometrical arrangements of grains in the disk to be experimented with. The disk scale-height and the inner and outer radii are free parameters for each of the disk components. Additionally, we fit for the total disk mass and the relative contribution by each component. The scale-height exponent and the radial density exponent are fixed at values 2.2 and 1.2 respectively. These values are recommended by \citet{whitney} based on the hydrostatic calculations of \citet{dalessio}. The overall disk parameters were the total mass of the disk and the ratio of this disk mass between the two disk components. For one source, the spiral features of the code were also required, where the free parameters were the pitch angle of the spiral, the summation parameter (that determines the width of the arms), the fraction of mass entrained in the arms and the radius where the spiral arms begin. The cavity parameters investigated were the cavity opening angles, the cavity density exponent, and the coefficient for cavity density distribution. The envelope free parameters were the centrifugal radius, the minimum and maximum envelope radii, and the envelope infall rate.

The dust types within the models are not varied throughout the fitting process. The dust type for the large-grains disk follows a dust distribution based on models from \citet{wood02}, with grain sizes ranging between 5$\mu$m and 1mm. The cavity and `small grains' disk share the same dust type, based on that of \citet{kim} for the diffuse ISM (with maximum grain sizes of 0.01$\mu$m). For the envelope grains, \citet{whit03} produced a size distribution of grains in agreement with $R_{v} \sim 4$, which is typical of dense regions of molecular clouds \citep{whittet01}. To include the effects of ices in the cooler outer regions of the envelope a layer of water ice was included as the outer 5\% of the grains' surface. This final percentage was decided upon after variation testing and comparison to polarisation observations, as described in \citet{whit03}. Polyaromatic-hydrocarbons (PAHs) and very small grains (VSGs) can also be accounted for in the code, resulting in emission that is visible in the SED and also has effects on the other observables. If PAH emission is recorded for a source in the literature, the dust files which are identical to the above files are used, with the exception that they do not include the grain population with sizes smaller than 200\AA{}. These are instead calculated for using the dust distribution of \citet{draine}. PAHs were not detected for the source studied in Paper I, so the results of including these PAHs on our simulated observables are published here for the first time. 

The interferometric observables are simulated in Fourier space. HOCHUNK-3D model images are generated at 1${\mu}m$ intervals within the MIDI N-band range, the fast-Fourier transform is taken of the images, and the specific u-v point or visibility according to the position angle and baseline of the observations is extracted, corresponding to the data point obtained with MIDI. By applying this to each image, a visibility profile with wavelength is obtained. In order to simulate the Q-band images, model images are generated at the required wavelength and convolved to the telescope's resolution. Azimuthally averaged radially profiles are then used to compare the model, source and PSF images. The SEDs were generated with IDL post-processing scripts \citep{whitney}, which could then be compared with photometric data points compiled from the literature.

\subsection{Fitting process}
 
The fitting process is the same as that of Paper I, with the SED constituting the starting point for the fitting, before the high-resolution datasets were considered. The Ulrich-type envelope and cavities successfully used in \citet{wit10} and \citet{wheel} were the starting geometry for the envelope and cavity specifications in this work. The fitting process starts with the simplest geometry, which consists of only of an envelope. If a fit could not be obtained, then the model geometry becomes progressively more complex by changing outflow cavity parameters and adding disks and eventually disk substructure.

When combining a variety of observations, the balance between the fits of those observations needs to be considered. When the SED alone is utilised, a common method of uncertainty quantification has been the chi-square minimisation. The SED Fitter tool of \citet{robsed}, which utilises a large grid of models and has been used to analyse the data of many protostars, employs this minimisation. Once a method is expanded past SED considerations, which as previously discussed is important to better constrain protostellar environments, the uncertainties in the fit parameters must be and have been considered differently. This is because different observations, at different wavelengths and using different techniques will all trace different scales of MYSO environments in contrasting ways. Few papers dedicated to MYSOs in the literature fit RT models to more than one set of observations, but the papers that are most similar to our approach have tackled this is different ways.
 
\citet{linz09} fit MIDI interferometric data and an SED for one MYSO. They used the aforementioned SED Fitter tool to constrain the fits on their SED and compared the model visibilities associated with a variety of acceptable models from that process in order to decide upon their best fitting model. A similar approach was taken in \citet{wit10}, where the SED Fitter was again used to provide some quantitative analysis of the SED fits, but the overall fit between the MIDI and SED was determined by eye. In \citet{wit11}, where one MIDI dataset and an SED were used, the overall fit was determined by eye. \citet{kath13} modelled the SED and 2MASS image profiles of the MYSO AFGL 2591 using a Monte Carlo radiative transfer code. The reduced combined chi-square was calculated for the line profiles and their SEDs, giving equal weight to each observation within the final chi-square. They studied the constraints on their parameter space by systematically varying a few key parameters and investigating how the chi-squared varied with each parameter.
 
\citet{boley13} determined their best fits among 2D Gaussian models fit to their MIDI data by performing a grid search over their parameter space and the chi-square minimisation. The uncertainties on the fitting parameters were found using a Monte Carlo approach. This involved generating synthetic observations, which were normally distributed about the mean visibility of their observations, and repeating this to optimise. This was a viable approach given that only one observable (their MIDI data) was being used in their work. In \citet{boley2016}, a variety of different models were applied to a combined dataset of MIDI observations, T-ReCS aperture masking data and K-band interferometric data for MYSO. Gaussian modelling was again used to fit their visibilities, and the chi-square minimisation (as in \citet{boley13}) was used to determine good fits; however a different fit found for the different wavelength datasets and no overall chi-square values were computed. Temperature-gradient disk models were also used to fit the K-band visibilities and the N-band correlated fluxes and a reduced chi-squared was calculated across the different observations; however, the weighting was not explicitly explained. Finally, radiative transfer models were employed via the use of the 2003 version of HO-CHUNK \citep{whit03}. They did not attempt to reproduce the SED throughout this modelling; only the K-band visibilities and N-band fluxes were modelled. No errors are formalised on their final parameters, and the chi-square was not utilised during this portion of their analysis.

In \citet{mengyao}, the model grid of \citet{zt} was used to fit the SEDs of a number of MYSOs. Following the method of \citet{deb17}, they performed the chi-square minimisation to find five best fitting models for each of the MYSOs. Five parameters alone were varied within the fitting process in order to investigate how the SED is affected by models corresponding to different evolutionary phases that occur within the turbulent core model \citep{tanmsf}.

The aforementioned works constitute those most similar to the work on this sample, but more observations are used here and all are fit simultaneously. Hardly any of these works calculated a global chi-square for different datasets, nor did they derive formal errorbars on their parameters when RT codes were used. This appears to be the inherent compromise of combining many datasets and using a complex modelling approach. Including multiple observations and a large model parameter space provides a much better picture of a source, but formalising the combined error using these methods becomes arduous. Calculating an overall chi-square and giving each observation equal weight, as has been done in some of the works mentioned in the previous subsection, was not determined to be an appropriate approach for the work on our sample. The MIDI spectrum and VISIR profiles have dozens of points covering a very small wavelength regime, while the SED covers a wavelength regime from one to thousands of micron with a (most frequently) small number of data points. This means that there will be an obvious bias within any chi-squared value calculated for the combined datasets, in favour of the 7-13$\mu$m and $\sim$20$\mu$m datasets. Another reason to not combine the chi-squared is that each observation traces very different scales so, similarly, the innermost and smallest geometrical features of the source, which are traced by the data-rich MIDI spectrum, will be unfairly focused on through the use of a combined chi-square. 

Overall, it can be concluded that while the chi-squared may not be an appropriate way to provide a blanket estimate of a fit when multiple observations that trace different scales are used, it can assist in the determination of which observables are key for tracing different physical characteristics of MYSOs. In order to comment on the uncertainties involved in the fitting of the wide parameter space provided by the RT code, the chi-squared was calculated for each observation. The chi-squared was calculated separately for each configuration of the MIDI observations and then averaged. The image profile chi-squares were calculated from the radial profiles up to the point where the background noise exceeded the rms. The SED chi-squared was calculated by convolving the model SED with filter fluxes corresponding to the available data points. While the SED was the starting point for the fitting (as in \citet{wit10, linz09}) the SED fitter tool is not used. As such, the fits of these models avoid the inherent biases of the Robitaille grid of models, which are discussed at length in \citet{rob17}.

Following the above considerations and investigations into the individual chi-squares, we find that the different parameters can be traced within certain ranges. The disk inner radius could be tightly constrained within 5$^{\circ}$, as could the cavity opening angles. Generally, significant changes were found in the MIDI fits for variations of the order 0.1 in scale-height in the two disk components. Similarly the cavity density exponent was constrained within $\pm$0.1 and notable changes were observed in the observables when the cavity density was varied by a factor of two. Some parameter,s such as total disk mass, outer disk radius, and envelope outer radius, are not optimally traced by IR observables and as a result are poorly constrained. On average, the disk outer radius could usually be constrained to within 500au, the disk mass could be constrained to an order of magnitude, and the envelope outer radius could be constrained within a factor of five. For sources that were particularly close or orientated such that the disk emission dominated or had a mass-dependent substructure such as a spiral, the disk mass could be better constrained, and we discuss these ranges on a source by source basis in the Appendix.

\section{Results}

  \begin{table*}
\caption{Short descriptions of the geometry inferred from the parameters of the final models of each MYSO.}              
\label{features}      
\centering         
\begin{tabular}{c c}          
\hline\hline                        
Name & Geometry \\ 
\hline                                   
G305.20+0.21 & large-grains disk component with sublimation radius three times larger than $R_{sub}$ ($R_{min}$=60au) \\
 & Low cavity densities \\
 \hline
W33A & Large and small grains disk components \\
\hline
NGC 2264 IRS1 & Flared disk with a mass heavy spiral that manifests as a small gap in the disk \\
\hline
S255 IRS3 & Flat large and small grains disk components, envelope a factor of ten smaller than the other sources \\
\hline
Mon R2 IRS2 & Inner hole \\
IRAS 17216-3801 & Cleared disk regions within a 100au radius, some dust still within this cavity ($R^{env}_{min}$<$R^{disk}_{min}$) \\
\hline
M8EIR & Flared disk with a cleared inner hole 20au in radius \\
\hline
AFGL 2136 & 125au radius inner hole \\
\hline                                             
\end{tabular}
\end{table*}

\begin{sidewaystable}
\caption{Parameters of the preferred models for the full sample of sources. PA is shorthand for position angle and is defined as east from north. Exf. stands for cavity density exponent \citep{whitney}, D1 SH is short for `disk 1 scale-height' and D2 SH is short for `disk 2 scale-height'.}        
\label{sampleparams}      
\centering            
\begin{adjustbox}{width=1\textwidth}
\begin{tabular}{c c c c c c c c c c c c c c c c c c c c}          
\hline\hline                        
Source & $M_\star$ & $L_\star$ & $i$ & $d$ & $R\SPSB{min}{env}$ & $R\SPSB{max}{env}$ & $R_{c}$ & $\dot{M}_{infall}$ & $\theta_{cav}$ & $n_{cav}$ & Exf & $M_{disk}$ & $R\SPSB{min}{disk}$ & $R\SPSB{max}{disk}$ & Dust Fraction & D1 SH & D2 SH & $A\SPSB{for}{v}$ & PA \\    
& ($M_{\odot}$) & ($L_{\odot}$) & ($^{\circ{}}$) & (kpc) & (au) & (au) & (au) & ($M_{\odot}yr^{-1}$) & ($^{\circ{}}$) & (gcm$^{-3}$) & & ($M_{\odot}$) & (au) & (au) & & & & & ($^{\circ{}}$) \\ 
\hline                                   
W33A & 25 & 49270 & 120 & 2.4 & 18 (R$_{sub}$) & 5$\times$ 10$^{5}$ & 500 & 7.4 $\times$ $10^{-4}$ & 20 & 1 $\times$ 10$^{-19}$ & 0 & 1 & 18 (R$_{sub}$) & 500 & 0.2 & 0.11 & 0.22 & 0 & 30 \\
G305.20+0.21 & 25 & 48500 & 35 & 4 & 60 & 5$\times$ 10$^{5}$ & 2000 & 7.5 $\times$ $10^{-4}$ & 12 & 8 $\times$ 10$^{-21}$ & 0 & 1 & 60 & 2000 & 1 & 0.1 & - & 1 & 55 \\
NGC 2264 IRS1 & 8 & 4200 & 15 & 0.74 & 4 (R$_{sub}$) & 5$\times$ 10$^{5}$ & 500 & 9 $\times$ $10^{-4}$ & 25 & 8 $\times$ 10$^{-21}$ & 0.25 & 0.3 & 4 (R$_{sub}$) & 500 & 0.2 & 0.4 & 0.8 & 25 & -18.5 \\
S255 IRS3 & 20 & 21550 & 120 & 1.8 & 12 (R$_{sub}$) & 1$\times$ 10$^{5}$ & 500 & 7.5 $\times$ $10^{-8}$ & 20 & 6 $\times$ 10$^{-19}$ & 0 & 1 & 12 (R$_{sub}$) & 500 & 0.2 & 0.1 & 0.1 & 70 & 67 \\
IRAS 17216-3801 & 38 & 172000 & 15 & 3.08 & 1 & 5$\times$ 10$^{5}$ & 1000 & 4 $\times$ $10^{-3}$ & 40 & 9 $\times$ 10$^{-21}$ & 1 & 1 & 100 & 1000 & 0.2 & 0.4 & 0.7 & 60 & 320 \\
Mon R2 IRS2 & 15 & 5500 & 130 & 0.84 & 20 & 5$\times$ 10$^{5}$ & 2000 & 9.5 $\times$ $10^{-4}$ & 30 & 3 $\times$ 10$^{-21}$ & 0 & 0.1 & 20 & 2000 & 0.2 & 0.4 & 0.8 & 25 & 190 \\
M8EIR & 13.5 & 12100 & 25 & 1.33 & 30 & 5$\times$ 10$^{5}$ & 2000 & 1 $\times$ $10^{-3}$ & 25 & 8 $\times$ 10$^{-21}$ & 0.25 & 0.1 & 30 & 2000 & 0.2 & 0.4 & 0.8 & 25 & 260 \\
AFGL 2136 & 20 & 151000 & 60 & 2.2 & 125 & 1$\times$ 10$^{5}$ & 2000 & 3 $\times$ $10^{-4}$ & 22.5 & 3 $\times$ 10$^{-19}$ & 0 & 20 & 125 & 2000 & 0.2 & 0.375 & 0.375 & 60 & -35 \\
\hline                                             
\end{tabular}
\end{adjustbox}
\end{sidewaystable} 

In this section, we present the results derived when applying the above methods to the sample of MYSOs. Table \ref{features} describes summarises the key features of each MYSO, while Table \ref{sampleparams} lists their specific parameters. An in-depth discussion of the fitting and features of each source is provided in Appendix A.

G305.20+0.21 is also included in the final sample but the results of its fitting have been previously published in Paper I. During the consideration of the full sample, we experimented with including the second disk component to see if improvements could be made to the fit of G305. Including the small grain disk component worsened the fit and, as such, the final results of Paper I are carried through to the comparative analysis made in Section 5. The final geometry for G305 is characterised by a disk with an inner radius of 60au, over three times the sublimation radius of the central protostar. Similar to the pilot object, G305.20+0.21 (see Paper I), all of the sources studied with our methodology appear to share the same base geometry - a central object surrounded by a disk, bipolar outflow cavities and an envelope. 

The central protostars of the models in this study cover a range of masses and luminosities. The protostellar masses were only varied from those obtained from the literature to improve the SED fits, as they had effects on the infall properties, and therefore the density distribution, of the envelopes. 

The envelope properties of the MYSOs were traced predominately by the longer wavelength emission of the SED and the 20$\mu$m emission, which is sensitive to the centrifugal radius (for closer sources) and the envelope infall rate. The larger the envelope infall rate, the larger the density within the infall radius (according to the Ulrich density distribution \citealt{ulrich}) and the larger the amount of material available for re-emission of the 20$\mu$m flux. The infall rate reaches values of $\sim$10$^{-3}M_{\odot}$yr$^{-1}$ for our sample. This is consistent with infall rates reported by \citet{mckeetan} for the turbulent core model. Accretion rates of this magnitude are reported in numerical work (e.g. \citealt{hosokawa}, \citealt{kuip11}), which is consistent assuming that envelope infall rates translate to dynamical mass accretion rates onto the star. The envelope outer radii are all similar across the sample, but our high-resolution data are not sensitive to such large scales. 
 All the MYSOs studied within our work have observables that can be fit with an Ulrich-type envelope, implying that all of the MYSOs are actively infalling and contain the subsequent density enhancements in the midplane that affect the 20$\mu$m emission. Follow-up kinematic studies could confirm this to be the case following methods such as those detailed in \citet{ohashi} (e.g. searching for inverse P Cygni line profiles in the sources' millimetre spectra). S255 IRS3 appears to be the exception in terms of envelope parameters, given its low envelope infall rate and smaller outer radius (by a factor of five). S255 is the only source confirmed to be episodically accreting so the fact that this source's envelope infall rate is lower is particularly interesting, as it could be related to the accretion activity and the fact that the observations in this paper correspond to an accretion lull. 

All the MYSOs in this sample have disks as a component of their final models. These disks are not toroids, which are 100s of solar masses \citep{bw}, but are objects that (while they could exist at larger radii) are present in the innermost ($\sim$100au) regions of the protostellar environment and within the accreting regions. The fact that all of the sample harbour a disk, lends support to the idea that massive forming stars accrete their mass through similar mechanisms to low-mass stars. 

The flaring and inner geometry of the disks had the greatest effects on the MIDI visibilities. For the majority of the model disks, the small grains disk component has a scale-height twice that of the large-grains disk. Such a ratio is in agreement with the hydrostatic calculations of \citet{dalessio} and models of observed images (\citet{cotera} and \citet{wood02}). IRAS 17216-3801 has a flaring factor very close to the value most commonly observed throughout the sample of 2 at 1.75, but the fact that a binary system is likely to be having dynamical effects on this disk could explain this. Two of the remaining disks are flat (scale-heights of the disks were 0.1) and one of them required no small grains to be present in the disk to successfully fit the datasets.

\citet{whitney} recommended that the distribution of the dust be split such that 20\% of the grains are in disk 1 (the large-grains disk component) and the remainder are in disk 2 (the small grain component). This was based on the expected dust settling timescales observed for low-mass source. Even though the exact nature of dust amalgamation within massive protostellar environments is yet to be accurately constrained, we find that this approximation suited all of the observations well, save for those of G305. The fact that no disk 2 was required to fit the MIDI visibilities for this source was discussed as a significant element that characterises this object in a possible transition phase in its disk evolution towards full disk dispersion, as discussed in Paper I.

Inner clearing, as found for the disk of G305.20+0.21, was also found to be present in the disks of M8EIR, AFGL 2136 IRS1 and Mon R2 IRS2. The degree of clearing varied between the four sources, ranging between 2-4 times the sublimation radius. NGC 2264 IRS1 displays a different form of substructure. 
The spiral parameters of the code were utilised, and we find that including a spiral with a loose pitch angle of 30$^{\circ}$, which contains 90\% of the mass of the disk and begins and ends at 30 and 60au, respectively, assisted in fitting the observables. These parameters essentially create substructure that manifests as a disk with a gap rather than an obvious spiral. NGC 2264 IRS1 is by far the closest MYSO of the sample and, as such, was traced at the smallest scales, so the fact that evidence is found for disk substructure only in the nearest of all sources in our sample is deemed consistent.

Of the disk parameters, one weaker constrained parameter is the disk mass. The only source where the disk mass was constrained to less than a factor of ten was NGC 2264 IRS1. This is likely due to the fact that it is near, that it has a close to face-on inclination, and that it has a mass-dependent substructure. 
The outer disk radius is only loosely constrained by the observables. A larger disk creates more millimetre emission, which can affect the longer wavelength regime of the SED and the maximum disk radius is also tied to the centrifugal radius, which affects the shape of the SED particularly at its 70$\mu$m peak and around 20$\mu$m. The imaging profiles also proved decent outer disk radii tracers, but mostly for the nearer sources. As such, the outer radius is still discussed but it is noted that this is not as constrained as the inner radius, which heavily affects the N-band emission in all cases. 

\section{Discussion}

We have performed RT modelling of a sample of eight MYSOs invoking a star-disk-envelope configuration. In the following section, we discuss common features and traits identified in the sample. Detailed discussions of the individual sources can be found in Appendix A. 

\subsection{Comparing the MYSOs of the sample}

Two factors are also defined to assist in the interpretation of the parameter space, resolving factor and flaring factor. The resolving factor was calculated to determine whether the resolving power of the MIDI observations affected the quantities derived. This was calculated by dividing the largest projected baseline \citep{vj07} used to observe each source by the distance to the source. The flaring factor is simply the scale-height of the `small grains' disk (disk 2) divided by the scale-height of the `large-grains' disk (disk 1). 

In order to make general comparisons between the sources in our sample, we calculate Pearson correlation coefficients. Correlations were inspected for basic parameters (bolometric luminosity, central object mass and distance) and the modelling parameters (cavity opening angle, disk flaring factor, envelope infall rate, cavity density, centrifugal radius and minimum and maximum disk radius). 
We do not calculate correlations for any parameters that were determined to be poorly constrained, namely disk mass and envelope radius. 

Before considering the correlations they were checked for bias. 
To check whether the specific MIDI observations affected the correlations found, correlations between the resolving factor and the parameter space were calculated. Strong correlations were detected for two characteristics - the distance and stellar mass, which seem consistent, as a very distant source needs to be more massive or luminous to be detectable, but if a source is close and observed with small baseline lengths it may be comparably resolved as a source that is distant but observed with the largest baselines. 
We find that no significant correlations are found between the inclination and the rest of the parameter space and that therefore the results found for the sample are not a result of fortuitous orientations. We conclude that any strong correlations beyond these above are tangible results from the parameter space. The vast majority of the correlation coefficients are <0.5. This implies that while the sample all share the same base geometry, diversity still exists within the group. We discuss some correlations of particular interest in the next subsection. 

   \begin{figure}
   \centering
   \includegraphics[width=90mm]{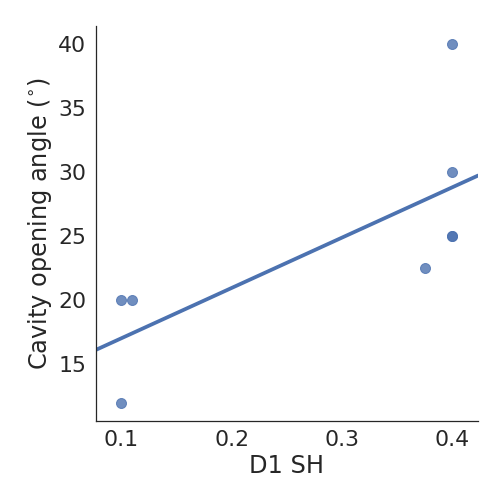}
   \caption{Relationship between the scale-height of the large-grains disk component (D1 SH) and the cavity opening angle.}
   \label{dshcav}
   \end{figure}
 
\subsubsection{Notable correlations}
 
  
   \begin{figure}
   \centering
   \includegraphics[width=90mm]{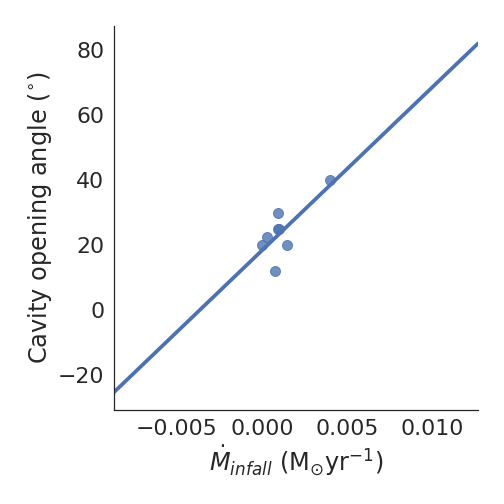}
   \caption{Relationship between envelope infall rate and cavity opening angle.}
   \label{infcav}
   \end{figure}  
  
A strong correlation is detected between the scale-height of disk 1 and the cavity opening angle (0.72) implying that the more flared the disk, the larger the cavity opening angle (Figure \ref{dshcav}). If the disk scale-height is large, the disk is more flared and a larger emitting surface will be visible. This same affect occurs if the cavity opening angle increases as this in turn increases the available area for irradiation by the protostar and therefore generates a larger emitting region. The correlation between disk 2 scale-height and cavity opening angle is moderate (0.62) because one model does not possess this second disk component. 

      \begin{figure}
   \centering
   \includegraphics[width=90mm]{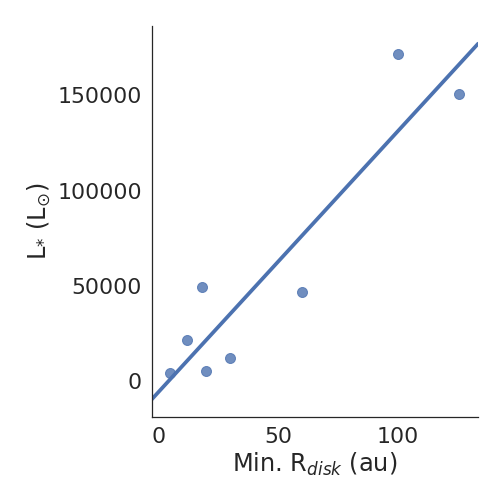}
   \caption{Relationship between the luminosity of the source and the minimum disk radius.}
   \label{lrmind}
   \end{figure}
   

  
   
A correlation of 0.75 is found between the envelope infall rate and cavity opening angle, implying that as the envelope infall rate increases, the cavity opening angle will get larger (Figure \ref{infcav}). 
The Ulrich geometry the envelope model uses is based on a rotating and infalling solution and results in a build of material at the equatorial plane. If the envelope infall rate is high, more material is moved from the outer envelope towards the midplane it becomes more concentrated while the regions closer to the poles of the envelope are less dense. The remaining envelope material would therefore be more vulnerable to the effects of any outflow activity from the central source, meaning that more material could be carved of the envelope, thereby widening the outflow cavity such that it has a larger opening angle, explaining why this trend is observed.
  
\begin{figure}
   \centering
   \includegraphics[width=90mm]{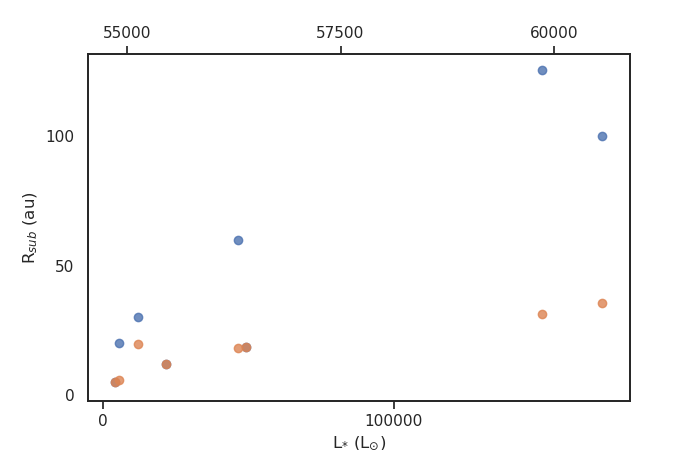}
   \caption{Relation between luminosity and sublimation radius (orange dots). For comparison, the blue dots denote the disk minimum radii as in Fig. 4.}
   \label{rminrsubcomp}
   \end{figure}

A strong correlation (0.92) exists between disk minimum radius and luminosity (Figure \ref{lrmind}). If considering straightforward dust sublimation this is to be expected, yet the minimum dust radius does not correspond to the actual dust sublimation for many sources. This is illustrated in Figure \ref{rminrsubcomp}. In Paper I causes of inner holes in disks were discussed at length and it was determined that the most likely causes for inner holes in massive stars are 1) photoevaporation and 2) the presence of binary/multiple companions. Of these two causes, photoevaporation as a mechanism seems to be in agreement with this trend. If a source is more luminous, then it will have a greater capacity to generate photons of high enough energies to dissociate the inner disk material and create cleared inner regions. Therefore, the large luminosity leading to a large inner hole makes sense as the star has another mechanism through which to destroy the disk. This comes with the caveat, however, that the star must not also be creating UV photons to the point where an ionising ultra-compact HII region has begun to form around the source, as that is not observed for any of these sources. A strong correlation does not exist between minimum envelope radius 
and luminosity, as the minimum envelope radius of the final model for IRAS 17216-3801 vastly differs from its minimum disk radius. If IRAS 17216-3801 is removed, the correlation between minimum envelope radius and luminosity is becomes strong and the correlations between luminosity and both disk and envelope minimum radius become 0.93.
 
\subsection{Substructure in MYSO disks}

In our sample of eight MYSOs, selected merely on the basis of completeness of observations, our analysis indicates that five sources have evidence for substructure in their disks. A disk with substructure in our terminology is contrasted with an azimuthally symmetric disk extending inwards down to the dust sublimation radius. For all except one, this takes the form of a cleared inner hole. The selection of objects based on completeness of IR spatial and SED data implies a number of things for our sample compared to massive stars in general. Firstly, the objects are relatively close by and only G305.20+0.21 lies at a distance greater than the typical 2-3kpc distance expected for MYSOs. Secondly, they are statistically expected to be at the lower end of the massive stellar mass distribution. Additionally, they possess somewhat wide outflow cones, such that a significant portion of the IR radiation from the inner disk regions can escape. 
Thus, finding objects with evidence of `transition disks' amongst this sample would not be totally unexpected. The presence of the inner holes tentatively suggests that within the idea of dissipation of the circumstellar environment, the equatorial disk structure is sensitive to advanced erosion prior to a complete disappearance of the overlying envelope. The likely causes of this inner clearing were discussed in Paper I, where it was suggested that the presence of a companion or photoevaporation are the most likely causes for the clearing in MYSOs. In order to confirm which of these mechanisms is responsible, high-resolution imaging with the ELT/METIS or a much denser filling of the uv-plane with VLTI/MATISSE could be employed to investigate dust sublimation and disk evaporation and how this relates to the physical presence of an inner disk rim.  

The remaining source, NGC 2264 IRS1, showed substructure in the form of a gap-like spiral feature. The fact that these substructures are present has interesting implications for the massive star formation process. To date, substructures have rarely been reported around massive protostars. \citet{sannawarp} detect a sub-Keplerian disk and jet system around the MYSO G023.01-00.41 and find that the CH$_3$OH emission at scales $\sim$2000au is significantly warped with respect to the plane. \citet{maudafgl2} detect a ring-like enhancement in the disk of AFGL 2136, whose size scale matches the inner rim in our final model (see Appendix A for further details). Various other disks (e.g. \citet{jilee18}) show irregular features as a result of fragmentation. 
The substructure that improves the fit of the data for NGC 2264 IRS1 manifests as a gap in the disk. Various physical processes have been suggested that result in a well-defined radial discontinuity within a disk with a significantly lower density or even devoid of material. 
The first is the presence of a forming planet. \citet{pinte} present one of the first confirmations of a planet in a gas and dust gap in the protostellar disk of the 2.4M$_{\odot}$ object HD 97048, but otherwise the presence of planets in these gaps has mostly been inferred. As discussed in the context of G305 in Paper I, planets are not observed around massive stars but whether they can never form or cease to exist as the protostar becomes more luminous with age has not been confirmed. An alternative method of gap creation is dust trapping. Dust traps \citep{whipple} occur when a local gas pressure maximum in the disk attracts and traps dust grains allowing clearing to occur in the disks and has also been proposed as a mechanism for observed gaps in low-mass disks (e.g. \citet{gonz17}). Recently \citet{marel} illustrate through physical-chemical modelling that the gaps observed in ALMA data of the disk of the Herbig Ae star HD 163296 can be reproduced from dust trapping alone, without the presence of a planetary body. Dust growth presents another avenue of gap formation whereby the amalgamation or pile-up of dust grains clears space in the disk (e.g. \citet{zhang15}, \citet{oku}). Instabilities, both large-scale \citep{lb} and gravitational \citep{ti}, have also been shown through modelling to be able to cause gaps. Among these processes, the ones that would operate in a disk around a massive star are probably grain growth and instabilities, provided that the young massive star accretes from a high density (and massive) disk. Toroids, which have been shown to create instabilities in works such as \citet{lb}, have been observed around massive forming stars (such as those studied in \citet{belt05}), presenting an avenue for this instability and therefore gap to occur. Gravitational instability is a more likely occurrence in massive systems where the central protostellar mass is higher and when multiple systems are present. \citet{price18} have shown through hydrodynamical simulations that spirals, shadows and asymmetrical structures can all be produced by a companion star on an inclined and eccentric orbit approaching periastron. Both the effects of binaries and dust growth take 10s of thousands of years to occur, and are therefore also better suited to an MYSO in a later phase of formation.

\section{Conclusions}

In this paper we have presented a multi-scale and multi-wavelength analysis of a sample of eight MYSOs with the aim to make direct comparisons of their characteristics. We derive physical parameters from the observational data by means of radiative transfer modelling, fitting simultaneously spatially resolved and SED data. The RT model consists of a disk-outflow-envelope morphology and we find that such a geometry provides reasonable fits to the data. Envelopes were best described by Ulrich-type density distributions, suggesting that they could be dynamically unstable and implying that they are actively infalling environments. 

Agreement exists between the characteristics found through this work and those found through previous studies. From the study of this sample it appears that the the $\sim$10$\mu$m interferometric data (probing $\lessapprox$100au scales) are most sensitive to the inner disk geometry of the sources. This could be determined thanks to the combination of the interferometric data with the $\sim$20$\mu$m single-dish data which provides additional constraint on cavity material. Variation is seen at the small scales ($\lessapprox$100au), in particular within the disk geometry. Substructure is detected for 75\% of the sources, that is the disks deviate from an azimuthally symmetric disk extending inwards down to the dust sublimation radius. The most common form of this substructure is an inner hole, implying some dispersal mechanism is present within the inner environments of the MYSOs in addition to dust sublimation. A strong trend observed between the minimum dust radius and the luminosity of the central source implies that photoevaporation may be a likely candidate for the observed clearing. The final source with substructure, NGC 2264 IRS1, displays a gap-like structure. Given that increasing the fraction of large dust grains in the disk improved the fit to the N-band data for this source, and the high multiplicity fraction for MYSOs \citep{sausagebinary}, we suggest that dust growth and the presence of a binary/multiple system are the most likely candidates for any substructure in this disk. All the inferred substructures have been previously observed in evolved low-mass protostellar disks. This implies that despite their shorter evolutionary timescales, massive disk evolution follows a similar path as low-mass disk evolution. Future work to detect multiplicity and further improve the view of these inner regions through spectrointerferometric, imaging and image reconstruction methods, will assist in confirming these substructures and distinguishing their origins. 





\begin{acknowledgements}
We thank the anonymous referee for their helpful pointers and Prof. Melvin Hoare and Prof. Steven Longmore for their discussions, all of which enabled the improvement of this work. We also thank the STFC for funding this PhD project. The modelling presented in this work was undertaken using ARC3, part of the High Performance Computing facilities at the University of Leeds, UK.  
\end{acknowledgements}

\bibliographystyle{aa}

\bibliography{first}

\begin{appendix} 

\section{In-depth presentations of each source}

This section is dedicated to each object. For each object we 1) provide an introduction on the object itself 2) discuss any interesting effects on each of the simulated scales when experimenting with the parameter space of the RT code, and 3) present the results and final best fitting model for each object.

G305.20+0.21 is also included in the final sample but the results of its fitting have been previously published in Paper I. During the consideration of the full sample, we experimented with including the second disk component to see if improvements could be made to the fit of G305. Including the small grain disk component worsened the fit, and as such the final results of Paper I are carried through to the comparative analysis made in Section 5. The final geometry for G305 is characterised by a disk with an inner radius of 60au, over three times the sublimation radius of the central protostar. 

              \begin{figure*}
   \centering
   \includegraphics[width=140mm]{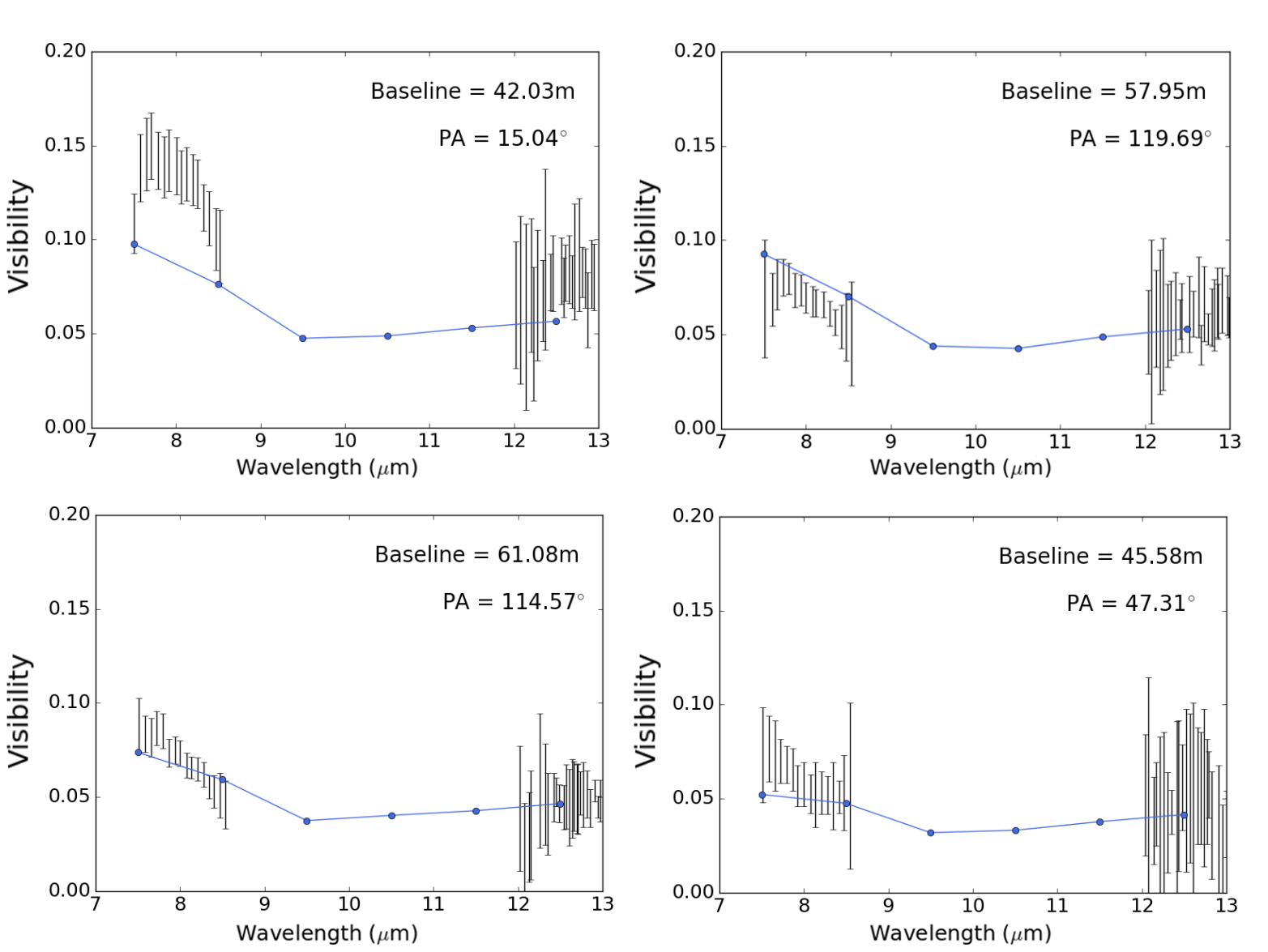}
   \caption{Observed visibilities for W33A for each configuration (black) with the simulated visibilities (coloured) for different models to show the effects different components on the simulated MIDI data. }
   \label{w33amidi}
   \end{figure*}
       \begin{figure}
   \centering
   \includegraphics[width=90mm]{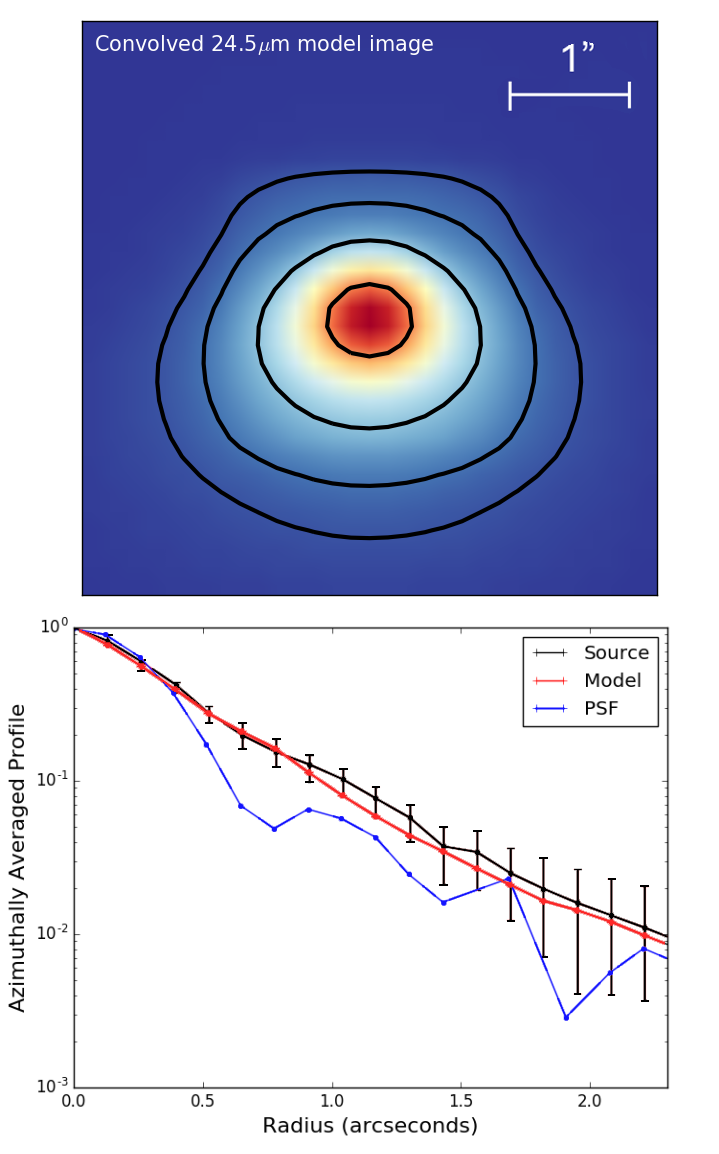}
   \caption{Convolved model image (top, contours represent 5, 10, 25 and 75\% of the peak flux) and the radial profile (bottom) of the best fitting model for the 2.4kpc distance. The original COMICS image of W33A can be found in \citet{wheel}.}
   \label{w33avis2}
   \end{figure}
    \begin{figure}
   \centering
   \includegraphics[width=90mm]{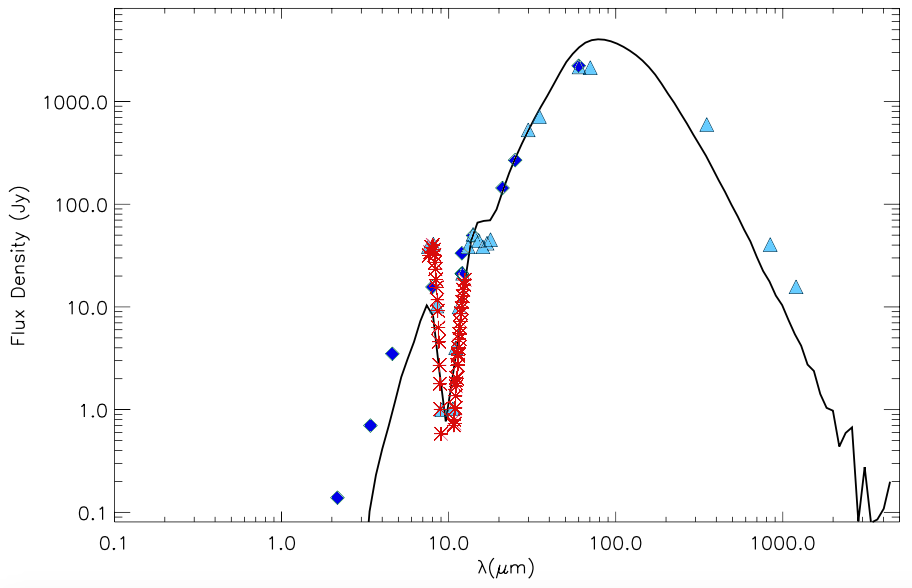}
   \caption{SED of the preferred final model for W33A. Data from the RMS survey are shown as diamonds, the MIDI spectrum is shown as red crosses and data points from \citet{wit10} are shown as triangles.}
   \label{w33ased}
   \end{figure}

\subsection{W33A}

 \citet{wit10} investigated the MYSO W33A by fitting an RT model to the SED and MIDI visibilities. They found that the source's MIDI data and SED could be reproduced by a model consisting of an envelope geometry with two bipolar outflow cavities and concluded that the majority of the mid-IR emission originates from the outflow cavity walls, in agreement with other works such as \citet{deb}. \citet{jilee13} attribute the CO bandhead emission of W33A to a 1-2au disk, while \citet{davies10} find a Br$\gamma$ jet perpendicular to a disk found in CO emission and absorption using NIR adaptive optics integral field spectroscopy. At longer wavelengths, two large continuum sources are observed within W3, the brighter of which is MM1 \citep{galvmad} or W33A. \citet{maudw33a} studied W33A with ALMA and SMA data to investigate whether a Keplerian disk may be present. They detected a change in position of the blue-shifted and red-shifted emission close to the peak of the continuum emission, which could be indicative of a rotating disk. They probed complex large-scale structure, with hints of a disk at smaller scales. The observations used in our study of the source consist of the MIDI observations from \citet{wit10}, the COMICS image presented in \citet{wheel} and an SED compiled from the RMS database and \citet{wit10}.

The luminosity of the source is listed as 3.4$\times$10$^{4}$L$_{\odot}$ by the RMS survey \citep{mottlum}. The kinematic distance to W33A was thought to be 3.8kpc \citep{faundez2004} but water maser measurements by \citet{immer} recalculate this as 2.4$\pm$0.2kpc. Our work therefore updates the mid-IR view of W33A substantially, updating it for the correct distance, adding to previous methodologies by including imaging data as an aspect of the fitting process and by using a more sophisticated radiative transfer code and due to the handling of the interferometric field of view (as discussed in Paper I). A test was done to see if the fit could be reproduced by simply reducing the luminosity of the central source according to the change in the distance. When done, the simulated visibilities for all the MIDI configurations increased between 8.5-10.5$\mu$m and the visibility at 7.5$\mu$m for configuration i), taking them outside of the observed errorbars and worsening the fits. This may be because the sublimation radius of the source will have become smaller, presenting a more compact component for emission resulting in increased visibilities. The 24.5$\mu$m density profile was also too broad and the silicate absorption feature of the SED became deeper and in violation of the observed MIDI spectra. 


   
\begin{figure}
   \centering
   \includegraphics[width=90mm]{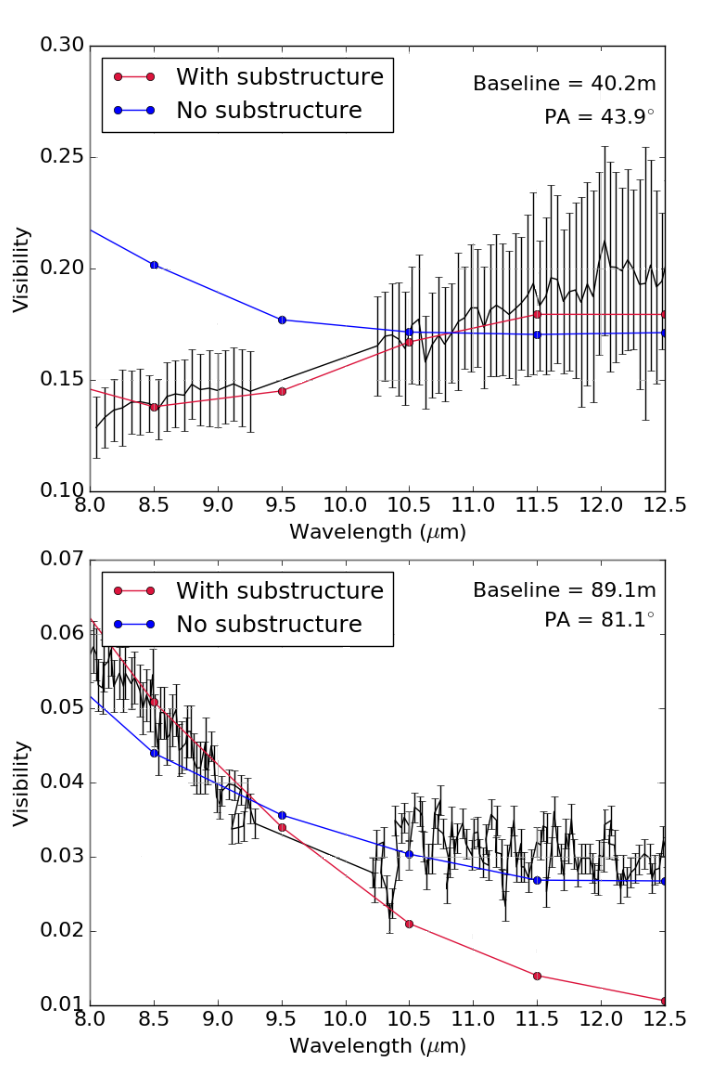}
   \caption{Observed visibilities of NGC 2264 IRS1 for each MIDI configuration (black) with the simulated visibilities for each model image (coloured).}
   \label{ngcvis}
   \end{figure} 

After determining that changing the distance alone could not provide a suitable fit to the dataset, recent literature, notably the existence of a potential disk \citep{maudw33a}, was taken into account when improving the fit. \citet{maudw33a} studied W33A with ALMA, tracing its wider environment. In their work they note the turbulent and messy nature of the source and image a stream of material that is spiral-like in nature on scales of $\sim$5000au. Although Keplerian signatures at the centre of the source are weak, they also state that while they do not detect a disk on scales of the spiral, a disk component could be viable at scales $\leq$500au. 

Including a disk within our models was experimented with, and improvements were found to the fits of the MIDI data. A disk can be included in the model with a maximum radius of 500au, in agreement with the suspected scales of \citet{maudw33a}. The disk can be made smaller than this without violating the fits but given the size of the protostellar environment and the suspected mass of the source, 500au is used as the final maximum disk value. This implies that the disk's emission from its inner regions is more important than the emission from its outer regions, which is consistent as the outer regions of a dense feature like a disk will be too cool to be traced by our high-resolution IR data. 
The disk is flat, with the large-grains disk component scale-height set to half that of the smaller grains disk. The model envelope has outflow cavities carved out at a 20$^{\circ}$ opening angle, twice that of \citet{wit10}, but with a similar cavity density. The envelope infall rates are identical between the two works. The inclination from \citet{wit10} for the 3.8kpc distance well satisfies all observables for the closer distance. The 
visibilities are displayed in Figure \ref{w33amidi}, the model 24.5$\mu$m image and the radial profiles are displayed in Figure \ref{w33avis2} and the SED is displayed in \ref{w33ased}. 

To summarise, W33A has demonstrated the importance of having the correct distance to MYSOs when deriving geometrical traits with the final model for the new distance characterised by the presence of a flat disk, which was not a part of the previous 3.8kpc model of \citet{wit10}.

      \begin{figure*}
   \centering
   \includegraphics[width=130mm]{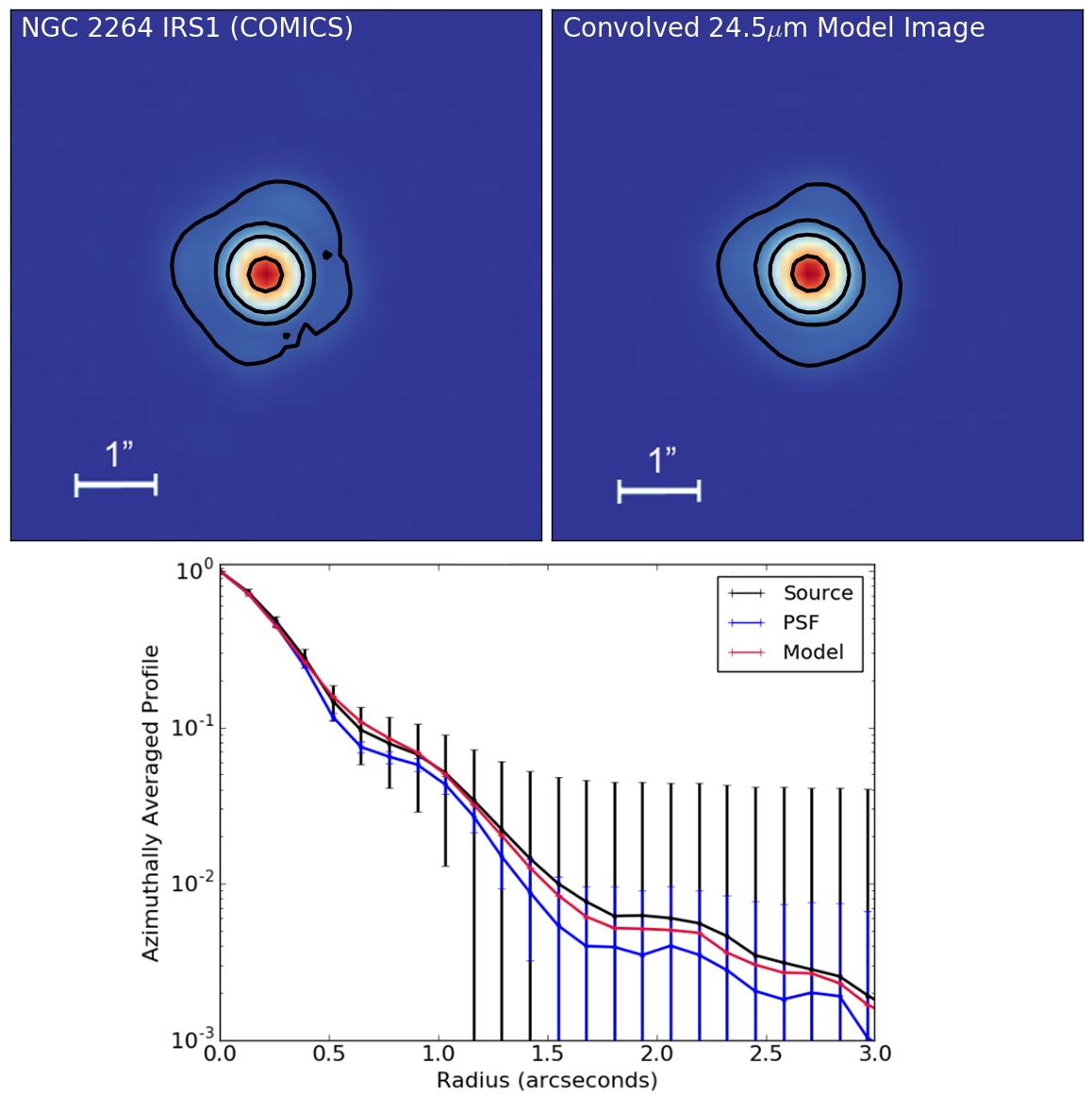}
   \caption{COMICS 24.5$\mu$m image (top left), convolved model image (top right) and subsequent radial profiles (bottom). The model image was convolved with the PSF of the observed object to accurately mimic the effects of the telescope specific to the observations. The contours in the images represent 5, 10, 25 and 75\% of the peak flux.}
   \label{ngcimgs}
   \end{figure*} 
         \begin{figure}[h!]
   \centering
   \includegraphics[width=90mm]{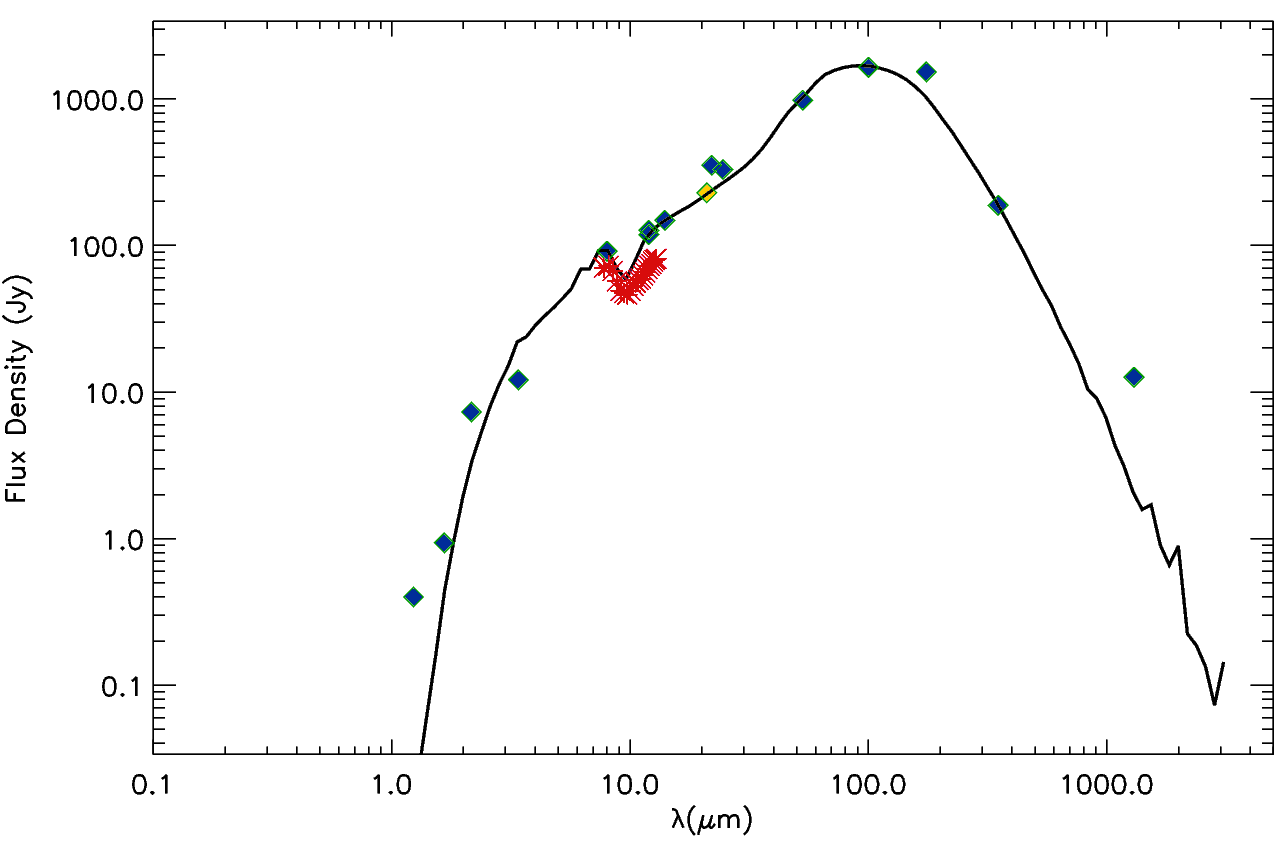}
   \caption{Model SED of the best-fitting model (black). Multi-wavelength flux measurements from the RMS are represented as blue diamonds, the yellow diamond represents the COMICS flux density and the fluxes corresponding to the MIDI visibilities are also shown in red.}
   \label{ngcsed}
   \end{figure}  
   
\subsection{NGC 2264 IRS1}

NGC 2264 IRS1 presented an interesting case for the fitting process due to its close distance of $\sim$740pc. During the fitting of other objects at more typical MYSO distances (2-4kpc) the COMICS profile was largely unaffected by the variation of disk parameters. In the case of NGC 2264 IRS1 however, the disk had profound effects on the profile and added another level of complexity to the fitting. Through the fitting it was determined that an Ulrich-type density distribution, as opposed to the power-law one of \citet{witcomics}, results in a model that produces a satisfactory fit to the datasets. Including the envelope is necessary to obtain a good fit to the longer wavelength data in the SED, which was not reproduced by previous work by \citet{grell}. The final model consists of a disk-outflow system, surrounded by this Ulrich envelope. The central protostar has a luminosity of 4.17$\times$10$^3$L$_{\odot}$ which is in good agreement with those found in the literature (\citet{harvey77}, \citet{sch85}). The stellar mass in the model is 9M$_{\odot}$, similar to the mass postulated by \citet{thompson} based on the source's B0-B2 spectral type. An inclination of 15$^{\circ}$ was able to successfully fit all the observables, which is consistent with the existence of an outflow in the line of sight found by \citet{ngcschrey}.  


Figure \ref{ngcvis} shows the MIDI data and two models which are discussed further later in this section. The two N-band configurations of NGC 2264 IRS1 trace significantly different areas of the protostellar environment, with large differences in both PA and baseline (an illustration is provided in Figure \ref{allmidiconfigs}). Both the MIDI configurations predominately trace disk material. The 40.2m baseline was characterised by its upward trend in visibility which proved challenging to fit. Often the issue was, again, that the model visibilities were too high between 7.5-9$\mu$m especially, indicating that the emitting regions at larger scales within the disk were also too compact and needed to be minimised. A near-perfect fit for Configuration ii) could be found when a disk was included with a 500au maximum radius consisting of 20\% large-grains (Model 1 in \citet{wood02}) and 80\% small grains (Galactic ISM grains as defined by \citet{kim}) but the fit for the 7-9$\mu$m region of the 40.2m baseline fit was poor in comparison. By adding substructure to the disk by using the 3D, spiral capabilities of the code, the fit for configuration i) could be improved (as described in Section 4). Including this `spiral' dramatically improves the fit of Configuration i) but worsens the fit of configuration ii) as can be seen in Figure \ref{ngcvis}. The overall change in the simulated visibilities for configuration ii) ($V^2\sim$0.02) is smaller than the improvement in visibility for configuration i) ($V^2\sim$0.08) so the model with substructure is referred to as the preferred fit. 

Given there is some evidence for PAH emission in the literature (e.g. \citet{pahs}), the inclusion of PAH emission in the code was experimented with. Adding PAHs/VSGs to all areas of the model protostellar environment increases the visibilities overall, vastly worsening the fit of configuration i). However, including the PAHs in just the cavity and envelope results in no change to the MIDI visibilities. This consecrates the fact that the MIDI data are dominated by the disk for this source. Ultimately a compromise is made by adding PAHs to the cavity and envelope only. This can be explained physically as PAHs constitute some of the smallest components of protostellar material and behave like very small dust grains. As a result, they could be vulnerable to the effects of photoevaporation. This means that they could be pushed from the inner environment into the envelope and cavity material at greater distances from the central protostar. Additionally, the PAHs in the further reaches of the envelope material would feel a weaker force from the stellar photons and would not be so easily removed from the environment. Therefore, the idea that PAHs may be present in outer regions of the protostellar environment but not the inner regions, may be consistent \citep{sieb}. The improvements noted by the inclusion of the spiral exist with or without the presence of PAHs.

Figure \ref{ngcimgs} shows the COMICS image, model 24.5$\mu$m image and the resulting radial profiles. The image is symmetrical and appears to show details of substructure, but \citet{witcomics} state that most of these are due to diffraction effects. Given its close distance, the 24.5$\mu$m emission of NGC 2264 IRS1 is heavily influenced by the disk included in the model. The close distance, inclination and the need for substructure within the model meant that the disk mass for NGC 2264 IRS1 could be more accurately constrained than other sources to an order of $\sim$0.1M$_{\odot}$. 
Changing the cavity density exponent, and therefore how evenly the cavity material was distributed, could assist in the fitting of the source. Increasing the exponent from the default value of 0 (constant density throughout the cavity) to 0.25 (meaning the mass was slightly weighted towards the bottom of the cavity, close to the protostar), reduced the amount of 24.5$\mu$m emission creating a better fit. This is consistent as more emitting material is at smaller radii within the protostellar environment, making the source appear less extended. A combination of the close-to-pole-on inclination and a foreground $A_v$ of 25 allowed sufficient fitting of the silicate absorption feature in the SED and is close to $A_v$ values quoted in the literature (\citet{thompson}, \citet{thomptok}). Changing the cavity exponent to satisfy the COMICS profile flattened the peak of the SED slightly, improving the fit. The SED fit is shown in Figure \ref{ngcsed}.

\begin{figure*}
   \centering
   \includegraphics[width=130mm]{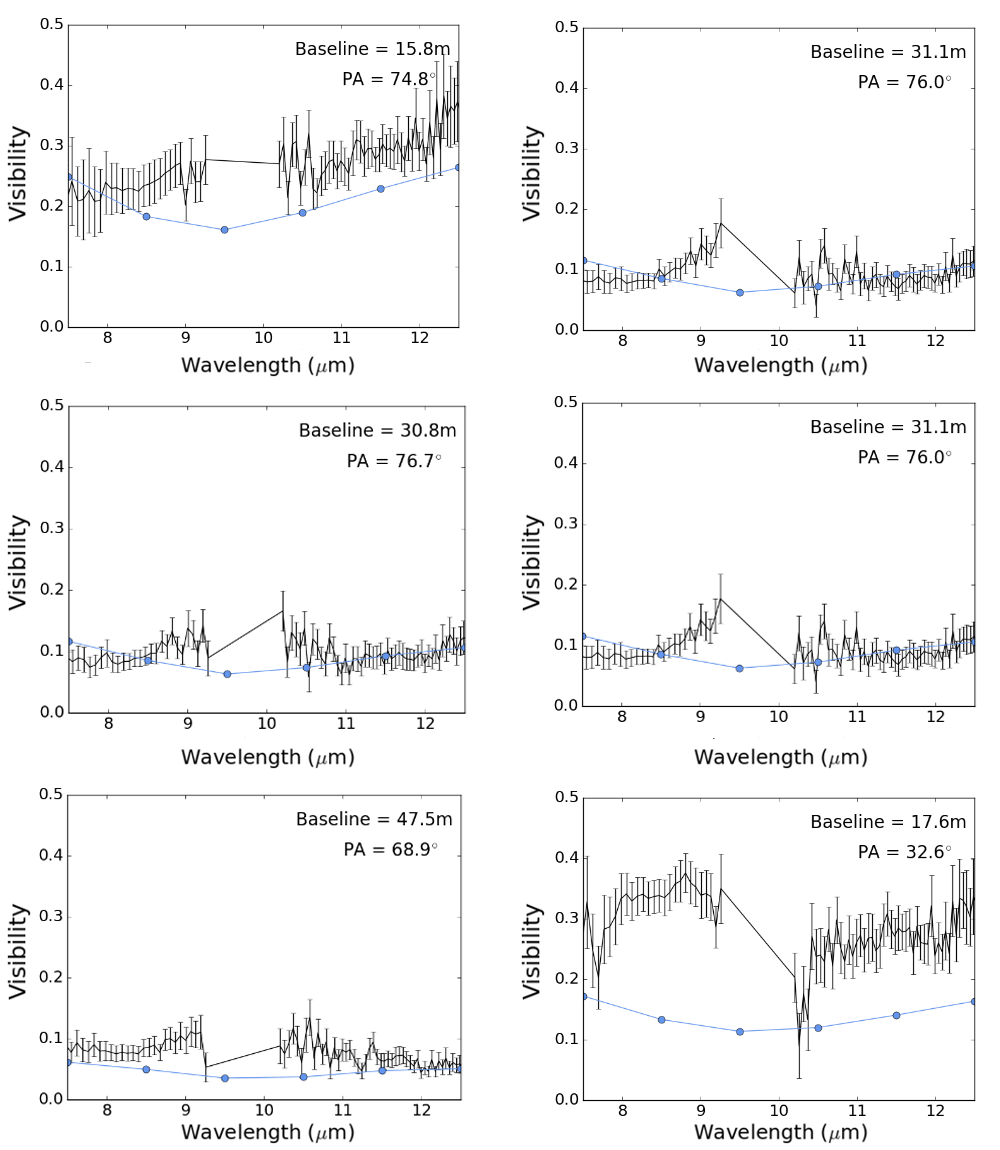}
   \caption{Observed visibilities for each configuration (black) with the simulated visibilities for each model image (coloured).}
   \label{s255vis}
   \end{figure*} 
         \begin{figure*}
   \centering
   \includegraphics[width=130mm]{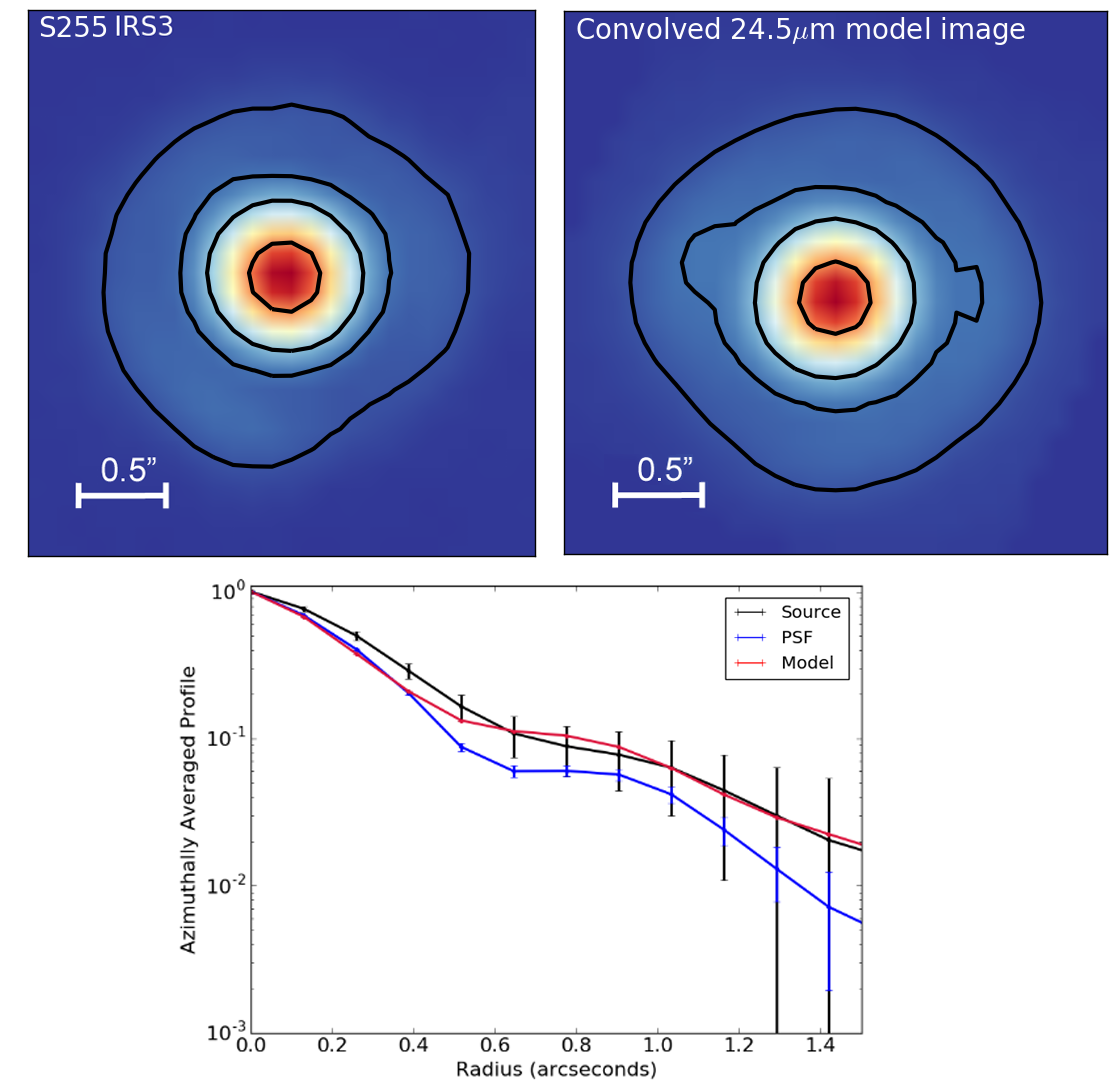}
   \caption{COMICS 24.5$\mu$m image (top left), convolved model image (top right) and subsequent radial profiles (bottom). The model image was convolved with the PSF of the observed object to accurately mimic the effects of the telescope specific to the observations. The contours in the images represent 5, 10, 25 and 75\% of the peak flux.}
   \label{s255imgs}
   \end{figure*}   

\subsection{S255 IRS3}

 S255 IRS3 has become of particular interest since an accretion-ejection event was observed around the source by \citet{car18}. A particularly bright 6.7GHz methanol flare was discussed in \citet{fuji} and a sub-millimetre flare was also found in \citet{zinchenko}. This flare was later tied to NIR and radio flares by \citet{car18} and these were determined to be the result of an accretion burst of order 10$^{-3}$M$_{\odot}$yr$^{-1}$. This was the first directly observed accretion burst around an MYSO, which has important implications as episodic accretion is a key component of the low-mass star formation process \citep{ep}. 

All the interferometric data for S255 IRS3 came from \citet{boley13}, with six configurations of data available for fitting. All the configurations, save one, trace very similar position angles of around 70$^{\circ}$. The remaining configuration traced a position angle of 32.6$^{\circ}$ and was low resolution due to its 17.6m baseline.

\begin{figure}[h!b]
   \centering
   \includegraphics[width=85mm]{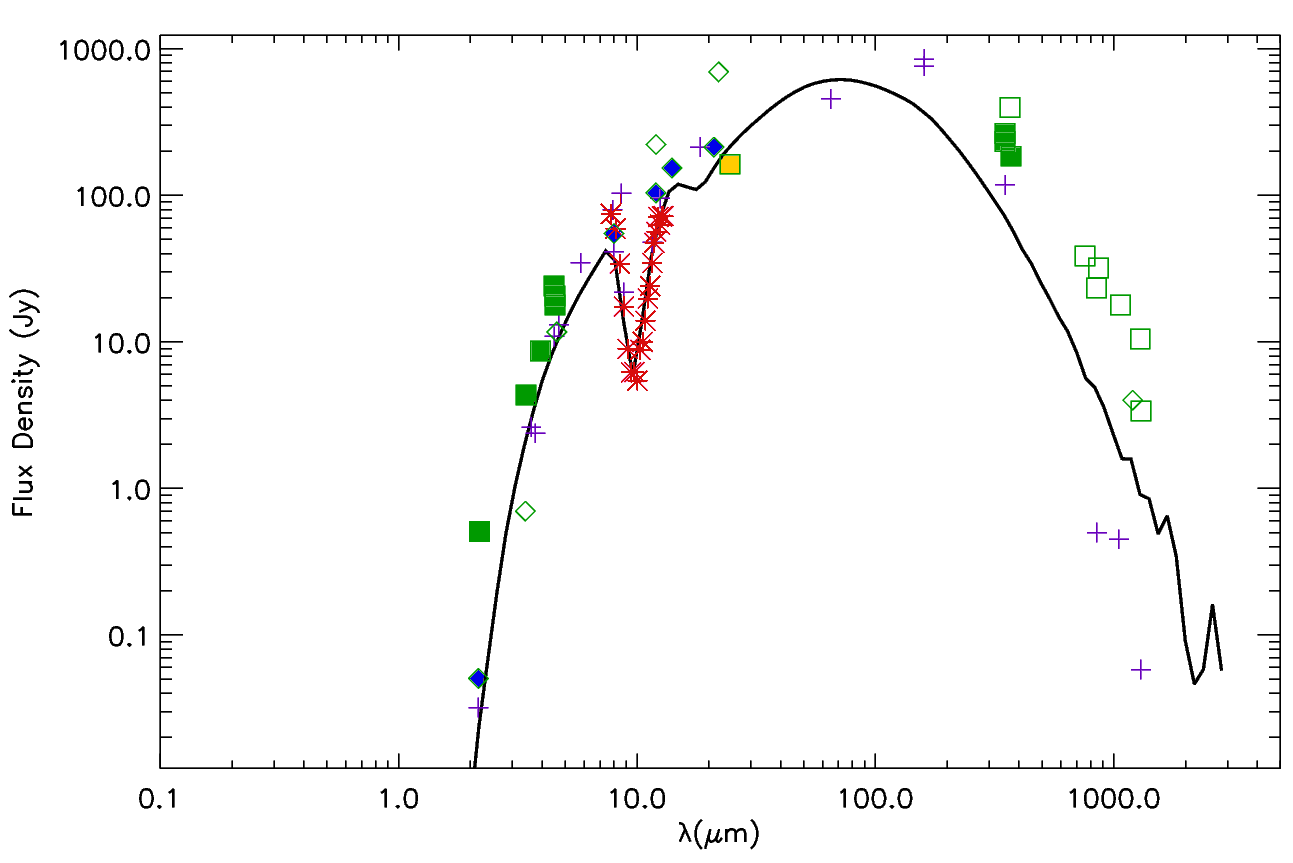}
   \caption{Model SED of the best-fitting model (black) for S255 IRS3. The different symbols correspond to different datasets, purple crosses are data from \citet{car18}, blue diamonds are fluxes from the RMS database, squares are data from \citet{witcomics} (with the COMICS flux in yellow) and the fluxes corresponding to the MIDI visibilities are also shown in red. Open symbols correspond to upper limits.}
   \label{s255sed}
   \end{figure}

The COMICS image from \citet{witcomics} is used in the fitting process. Both IRS1 and IRS3 are visible in the COMICS image. In order to exclude scales that may be influenced by the emission of IRS1 we fit the inner 3" of the source as opposed to the whole profile. All of the fitted data were taken at a time that appears to be an accretion lull for S255 IRS3, as the fluxes from the SED (which were taken at similar times to the MIDI and COMICS data) are in much better agreement with the non-burst fluxes from \citet{car18} than the burst fluxes. Fluxes from the RMS survey and \citet{car18} were incorporated (blue diamonds in Figure \ref{s255sed}), alongside additional fluxes from \citet{witcomics} (green crosses in Figure \ref{s255sed}) and the MIDI flux spectrum. 

The plots for the final model are shown in Figures \ref{s255vis} - \ref{s255sed}. The final luminosity for the central source in the model is 2.15 $\times$ 10$^{4}$L$_{\odot}$, which is in excellent agreement with that observed by \citet{s255l} of $\sim$2.4$\times$ 10$^{4}$L$_{\odot}$. The overall model is a disk and outflow system, in agreement with works such as \citet{zinchenko}, who studied the source with SMA and IRAM-30m continuum and spectral lines and find signatures of rotation and attribute this to a disk-like structure. 
The presence of the disk also agrees with the detection of the accretion event; as for low-mass stars, such events occur as a result of accretion from a disk surrounding the protostar. The fact that the envelope is rotating and infalling could be linked to the fact that the source has displayed an accretion event. Infall from the wider environment is directly linked to disk accretion in the turbulent core model \citep{tanmsf}. With a centrifugal radius of 500au, as opposed to 2000au, more material is fed closer to the central star, providing a larger reservoir of easily accessible material for accretion. This need not be a constant flow of material, meaning that the envelope infall rate and disk accretion rate can fluctuate throughout the life of the stellar system, allowing for low accretion rates. A flat disk with no inner clearing allows a good fit to be obtained for the MIDI visibilities of S255 IRS3. Two disk components were included in the successful model, a small grains disk and a large-grains disk, both with the same scale-height. Configuration vi) is the worst fit of the set. When consulting Figure 1 of \citet{car18}, we can see that this configuration is still likely to be tracing cavity material, as illustrated by Figure \ref{allmidiconfigs}. The simulated visibilities are lower than those observed implying that the smooth cavity surface presents too large an area for emission to match the data. This could be explained by the fact that the cavity material is clumpy and irregular \citep{car18}. The cavity density of S255 IRS3 is of order 10$^{-19}$gcm$^{-3}$, which is comparable to that of W33A.

      \begin{figure*}[h!]
   \centering
   \includegraphics[width=135mm]{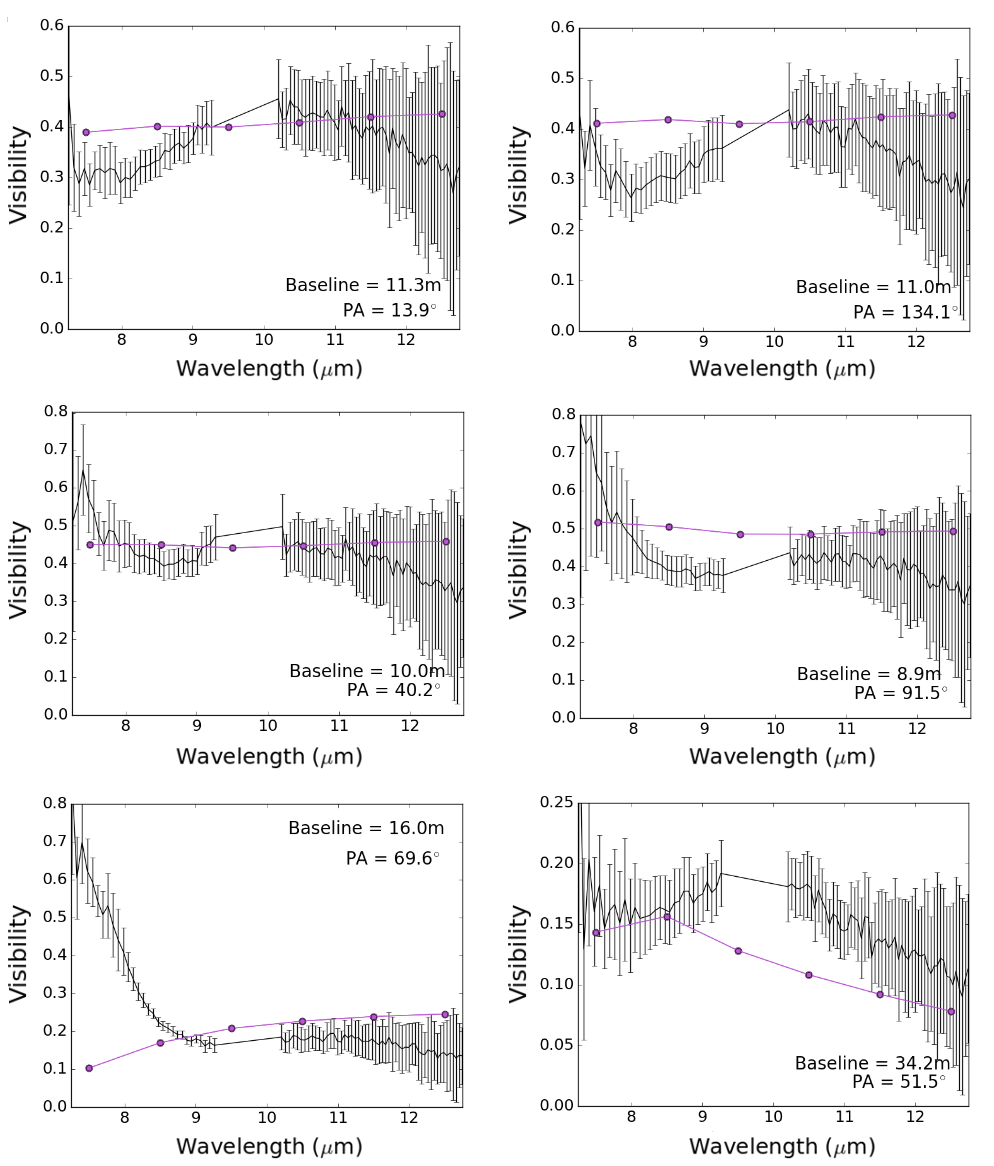}
   \caption{Observed visibilities of IRAS 17216-3801 for each configuration (black) with the simulated visibilities for each model image (coloured) (1).}
   \label{irasvis}
   \end{figure*} 
   
An inclination of 120$^{\circ}$ fit all the observables, which is consistent with the close-to-edge-on orientation suspected by \citet{boley13}. 
The foreground extinction towards the source is large ($A_{v}\sim$70) similar to the findings of \citet{car18}. Given the close to edge-on inclination, a disk mass from 0.1-10M$_{\odot}$ could be present and satisfy the observables which is poorly constrained due to the limitations of the IR as a disk mass tracer. A compromise was found for the longer wavelength regime of the SED between the data of \citet{car18} and the data from \citet{witcomics} (that they do not list as upper limits). 

Varying the disk parameter space has negligible effects on the COMICS profile, save for the disk outer radius and the centrifugal radius of the envelope, which is attributed to the sources inclination and distance. It was found that the 2000au maximum disk radius and centrifugal radius used for G305 resulted in too much 24.5$\mu$m emission to successfully fit the radial profile, implying that the disk must be less extended. Reducing these values to 500au allowed a good fit for the COMICS profile. 



\begin{figure}[h!]
   \centering
   \includegraphics[width=75mm]{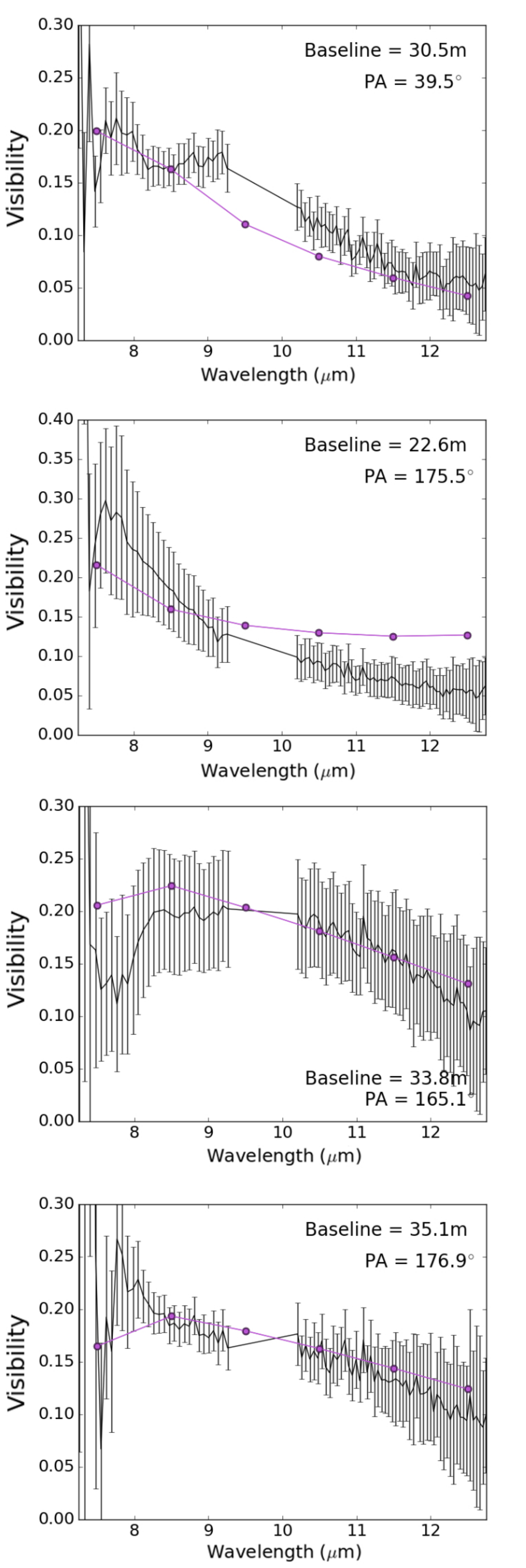}
   \caption{Observed visibilities of IRAS 17216-3801 for each configuration (black) with the simulated visibilities for each model image (coloured) (2).}
   \label{irasvis2}
   \end{figure}

The wider environment of S255 IRS3 differs compared to the other MYSOs in this sample, with an outer envelope radius that is five times smaller than the other sources and an envelope infall rate that is significantly lower. Raising the infall rate to a value comparable to those found for the other sources ($\sim$10$^{-4}$M$_{\odot}$yr$^{-1}$) increases the peak of the SED significantly. While this appears to improve the fit to the SED, the fluxes that are better fit are almost certainly detecting flux from both IRS1 and IRS3 and as a result this envelope infall rate is likely to be an overestimation of what is actually present. A fit with the lower envelope infall rate that fits the smaller aperture $\sim$70$\mu$m point better is preferred. High-resolution, small aperture observations at $\sim$100$\mu$m could allow separate SEDs for the two MYSOs to be obtained and for this to be better investigated. While obtaining and analysing these longer wavelength data are outside the scope of this work, increasing the envelope infall rate also lowers all the simulated MIDI visibilities that only trace IRS3, not IRS1. Using an envelope infall rate of the same order of magnitude as W33A and NGC 2264 IRS1 vastly worsens the fits for all configurations, further showing the need for a lower envelope infall rate. 

\subsection{IRAS 17216-3801}


\citet{kraus17} resolved IRAS 17216-3801 into a close binary using spectro-interferometric data from AMBER, CRIRES and GRAVITY. They find that the two stars are separated by $\sim$170au and that not only is there a disk surrounding the pair of stars but disks surrounding each individual object as well. HO-CHUNK3D does not provide the capabilities for modelling multiple systems but does present the opportunity to investigate the circumbinary material of IRAS 17216-3801. A single central object is included in the model, whose luminosity was obtained through the fitting of the SED as per the other sources in this work. 

          \begin{figure*}
   \centering
   \includegraphics[width=130mm]{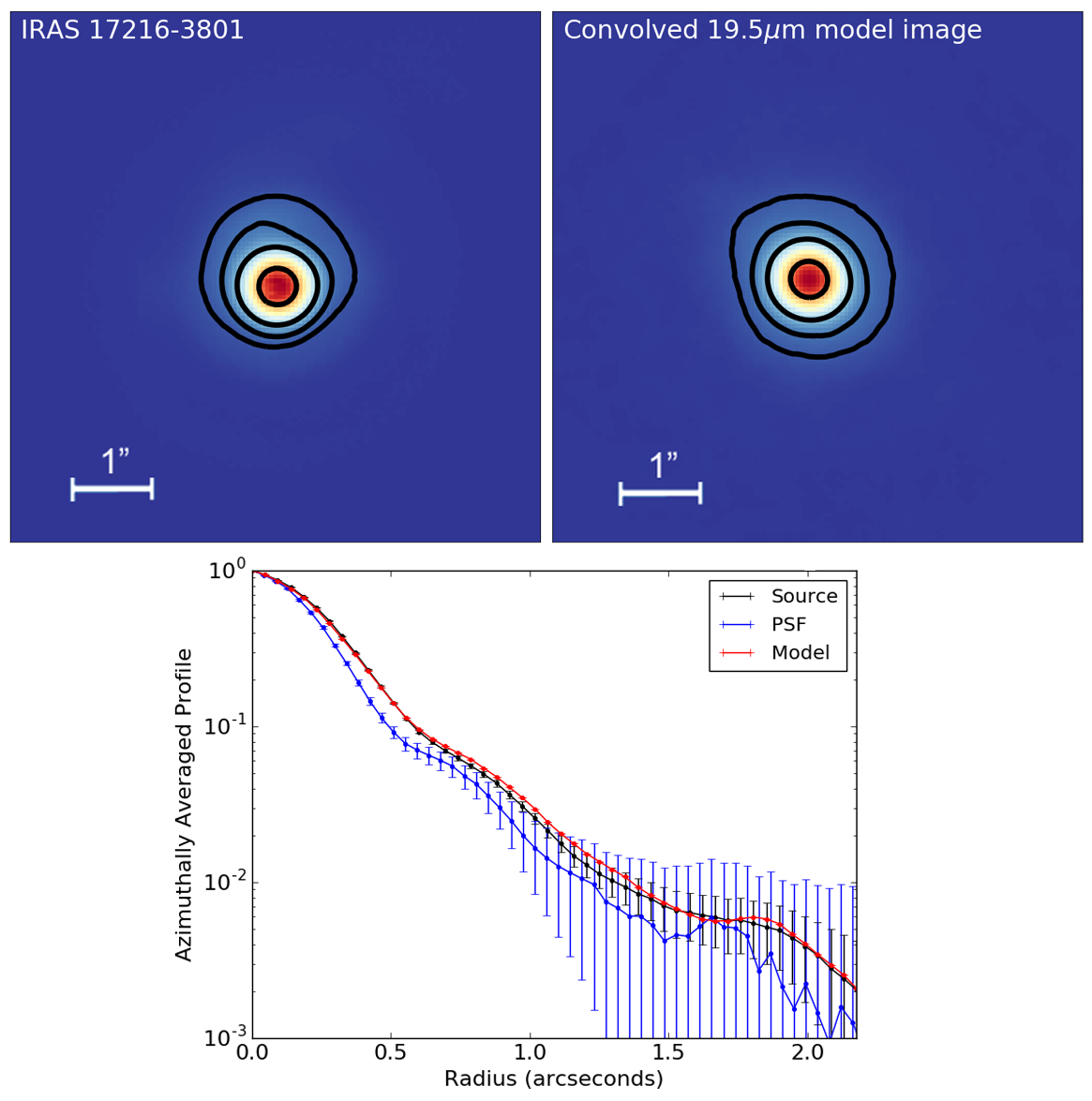}
   \caption{VISIR 19.5$\mu$m image (top left), convolved model image (top right) and subsequent radial profiles (bottom). The model image was convolved with the PSF of the observed object to accurately mimic the effects of the telescope specific to the observations. The contours in the images represent 5, 10, 25 and 75\% of the peak flux.}
   \label{irasimgs}
   \end{figure*} 

Surrounding this central source is a disk-outflow-envelope system. The disk scale-heights are very similar to the defaults suggested by \citet{whitney}, with the small grains disk being slightly flatter. The inner disk radius is 100au, which is comparable to the estimate of \citet{kraus17} for their circumbinary disk. The visibilities are shown in Figures \ref{irasvis} and \ref{irasvis2}. IRAS 17216-3801 has the largest MIDI dataset of our sample with ten configurations. The position angles of this dataset range from approximately 14-175$^{\circ}$ but all are fairly low resolution with baselines of 35m or shorter. This means that most of the MIDI configurations are sensitive only to the circumbinary material. As illustrated in Figure \ref{allmidiconfigs}, configurations viii), ix) and x) are the only three with PAs suitable for tracing the circumprimary disk and only configuration viii) has the potential to trace the circumsecondary disk. The baselines of configurations viii), ix) and x) are some of the largest for this source; configuration viii) can probe scales of $\sim$20-35mas, configuration ix) traces scales between $\sim$20-50mas and configuration x) traces scales between 19-33mas. As such these configurations not only have the PA to trace the circumstellar disks, but also the required resolution. Configurations vi) and vii) have comparable resolving power to configurations viii), ix) and x) but their PAs mean that they are more likely to be tracing any material within the cavity carved out by the binary. The baselines of the remaining configurations are short, meaning that they would all be tracing the suspected circumbinary disk material. The average visibility across each dataset varies considerably, and no trend is observed between the average visibility of the configurations that may be tracing circumstellar disk material and those that will not. It was found that changing the minimum envelope radius to values smaller than 100au improved the fits of configurations viii), ix) and x) by lowering the simulated visibilities across each configuration. As previously mentioned each of these configurations has the potential to be tracing circumstellar disk material, so the fact that including additional dust within the circumbinary disk improves the fits to these configurations appears consistent. If the configurations are not tracing circumstellar disk material, having a smaller minimum envelope radius than the minimum disk radius implies that the binary stars cannot have destroyed the dust in the inner hole nor accreted it into their binary disks. 




In order to satisfy the VISIR data (Figure \ref{irasimgs}) a low inclination was required (15$^{\circ}$) and a cavity density exponent of 1. This is because the VISIR profile is barely resolved. A density exponent of 1 as opposed to 0 (the default value) means that the density in the cavity decreases linearly with radius instead of remaining at a constant value. The innermost part of the cavity will then be brighter than its outer regions, reducing the extended emission and lowering the $\sim$20$\mu$m flux at larger radii, improving the fit. Similarly, close to pole-on inclinations reduce the extended emission as only one of the cavity lobes is directly visible in the line of sight. 

\begin{figure}
   \centering
   \includegraphics[width=90mm]{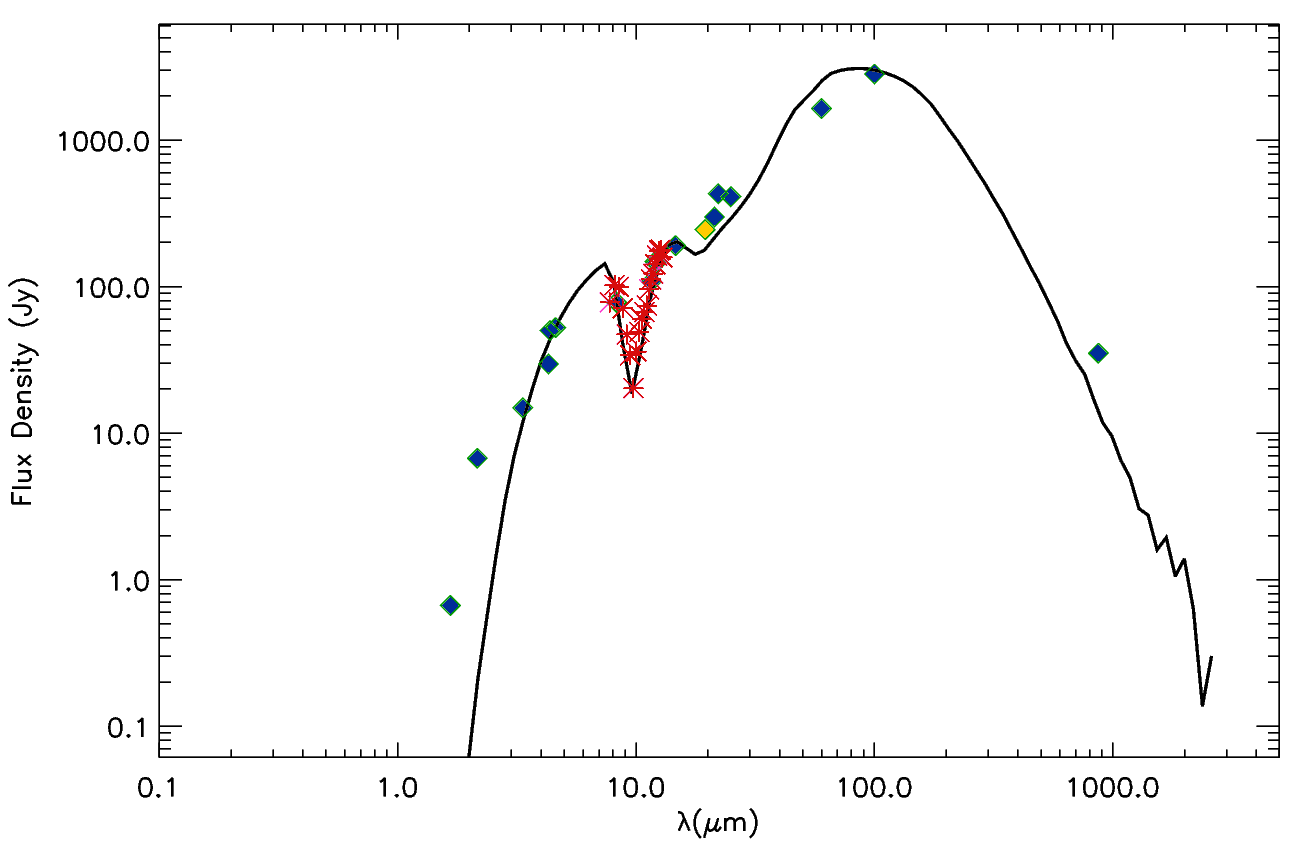}
   \caption{Model SED of the best-fitting model (black) of IRAS 17216-3801. Multi-wavelength flux measurements from the literature are represented as blue diamonds, the yellow diamond represents the VISIR flux density and the fluxes corresponding to the MIDI visibilities are also shown in red.}
   \label{irassed}
   \end{figure}

The SED is shown in Figure \ref{irassed}. The ATLASGAL 870$\mu$m point proved difficult to fit. The aperture of this measurement was 19.2" but the region is fairly uncrowded with no obvious sources contaminating such a measurement. In an attempt to improve the fit to this data point, disks of bigger radius were trialled. The larger the extent of the disk, the more cold emission at sub-millimetre and millimetre wavelengths will be present. A 4000au maximum radius disk showed small improvements in the millimetre of the regions but more substantial changes in the 20$\mu$m region of the SED, making the dip-like feature between 10-20$\mu$m flatten. This increase in radius caused all the simulated MIDI visibilities to decrease, worsening some fits and but improving the fit of configuration ix). The fit of the VISIR profile was also violated as the model radial profile displayed more extended emission than the observed profile. This occurred for any disks larger than 2000au so this was the final value chosen for the disk to preserve the quality of the VISIR fit. Making the envelope radius larger improved the fit to the longer wavelength end of the SED by shifting the SED to longer wavelengths, but this worsened the fit of the 20$\mu$m region of the SED.
   
The mass of the central source was also found to affect the shape of 10-20$\mu$m region of the SED. For example, if a 20 solar mass central source is used this 10-20$\mu$m region of the SED is flatter. 
This is consistent as the density distribution of the Ulrich envelope is proportional to $M_{*}^{1/2}$ \citep{ter}, meaning that as the stellar mass increases, the density will decrease within the infall radius and reduce the amount of 20$\mu$m flux.
Configuration v) proved the most difficult to fit. This occurred despite the fact that it is one of the lowest-resolution datasets and is likely to be tracing material within the circumbinary disk. A large (nearly eight-fold) upturn is present in the visibilities between 7 and 9$\mu$m, which is often a hallmark feature of a compact component like a disk. To test whether the very large increase in visibility could be due to the presence of the inner rim specifically, the inner disk radius was changed to the disk to 134au, which is the scale traced by this configuration at 7.5$\mu$m. A minor upturn in the visibilities was observed but the fits for the other, higher-resolution configurations worsened. When the inner radius of the disk was varied between 100-150au the visibilities changed marginally but the overall goodness of fit changed negligibly.

The upturn in configuration v) could be due to the binary system interacting with the circumbinary disk. Small-scale features, such as a puffed-up inner rim and disk-spiral, were tested but also present marginal changes, mostly likely due to the object's distance. A very specific local substructure must be present at the scales traced by this baseline alone and not the others of similar PA/baseline to induce the observed increase in visibility. Various theoretical works have shown that binary or multiple stellar systems can result in the formation of such a small-scale substructure in protostellar disks.  \citet{nelson} used the SPH code VINE to model the circumbinary disk of the low-mass source GG Tau A. They found that the binary system `stirs' the material in the disk, leading to the formation of spiral structures, which extend towards the centre within the inner radius of the disk as streams in the earliest stages. Within the disk itself, they see significant areas of over-density. \citet{thun} find that the inner rims of the circumbinary disks are often unstable, clumpy and recessing from their hydrodynamical modelling. \citet{desai} focus on the study of unequal mass binaries in particular using a 3D hydrodynamics code CHYMERA. Here they find that gravitational instabilities are induced by the binary interaction, with the smaller of the companions creating a density wave at the earliest stages and at later stages also see the generation of spirals. Observationally, \citet{tobin} detect spiral structures surrounding a three-star system. Although low mass, two of the stars have a separation comparable to that of IRAS 17216-3801 of $\sim$180au. These authors postulate that this distant star is the cause of disk fragmentation in the system and that the disk could have recently undergone a gravitational instability, inducing the formation of one or two of the companion stars. While all these cases focus on low-mass stars, the potential a binary system has to create gravitational instabilities and affect the stability and geometry of its circumbinary disk will be increased if those stars have higher mass. As such, the likelihood of over-densities that could create the compact feature required to fit this configuration is high and could therefore explain the poor fitting of these data.
   
\begin{figure*}
   \centering
   \includegraphics[width=140mm]{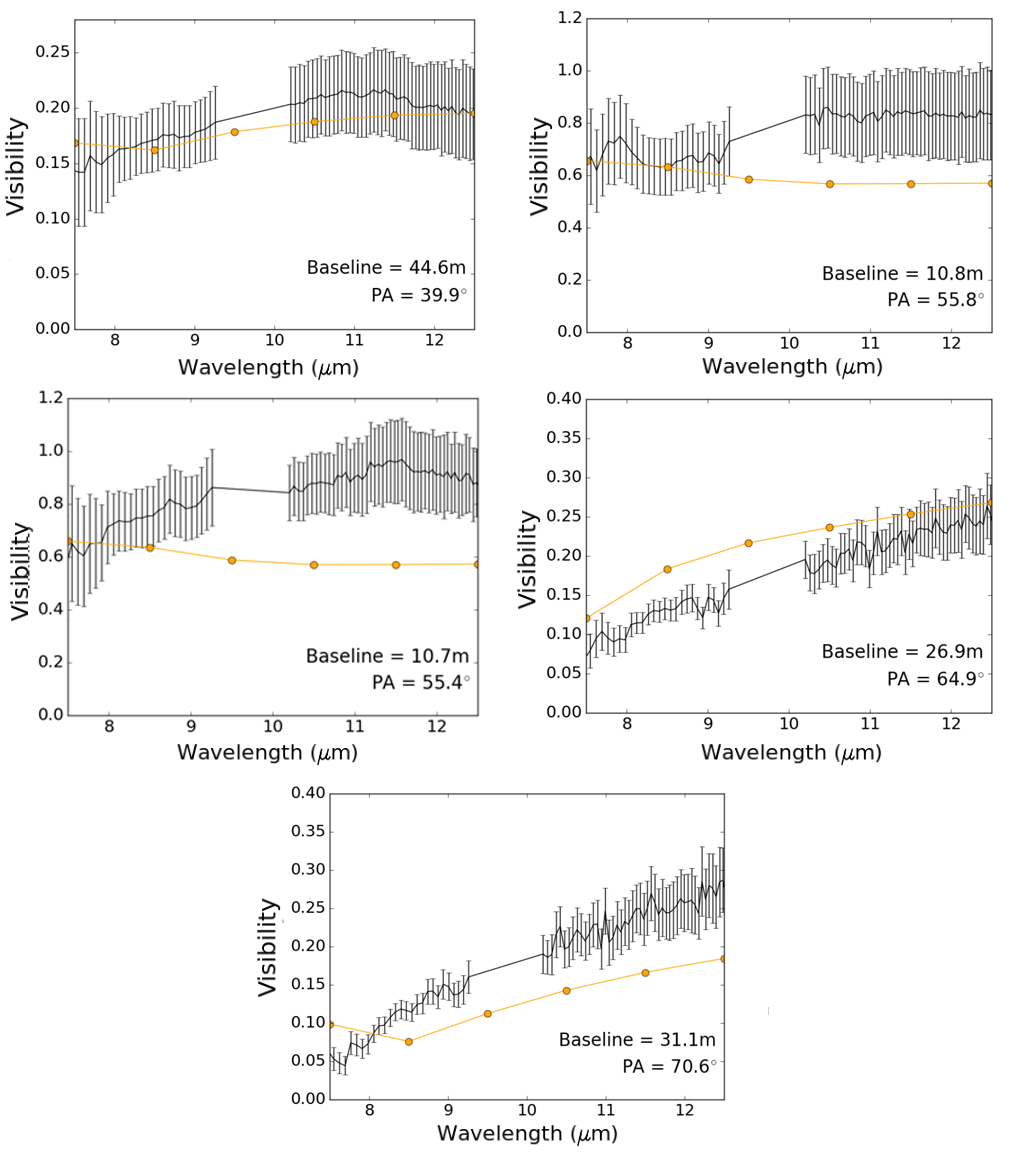}
   \caption{Observed visibilities for each configuration (black) with the simulated visibilities for each model image (coloured) for Mon R2 IRS2.}
   \label{monvis}
   \end{figure*} 
 
      \begin{figure*}
   \centering
   \includegraphics[width=130mm]{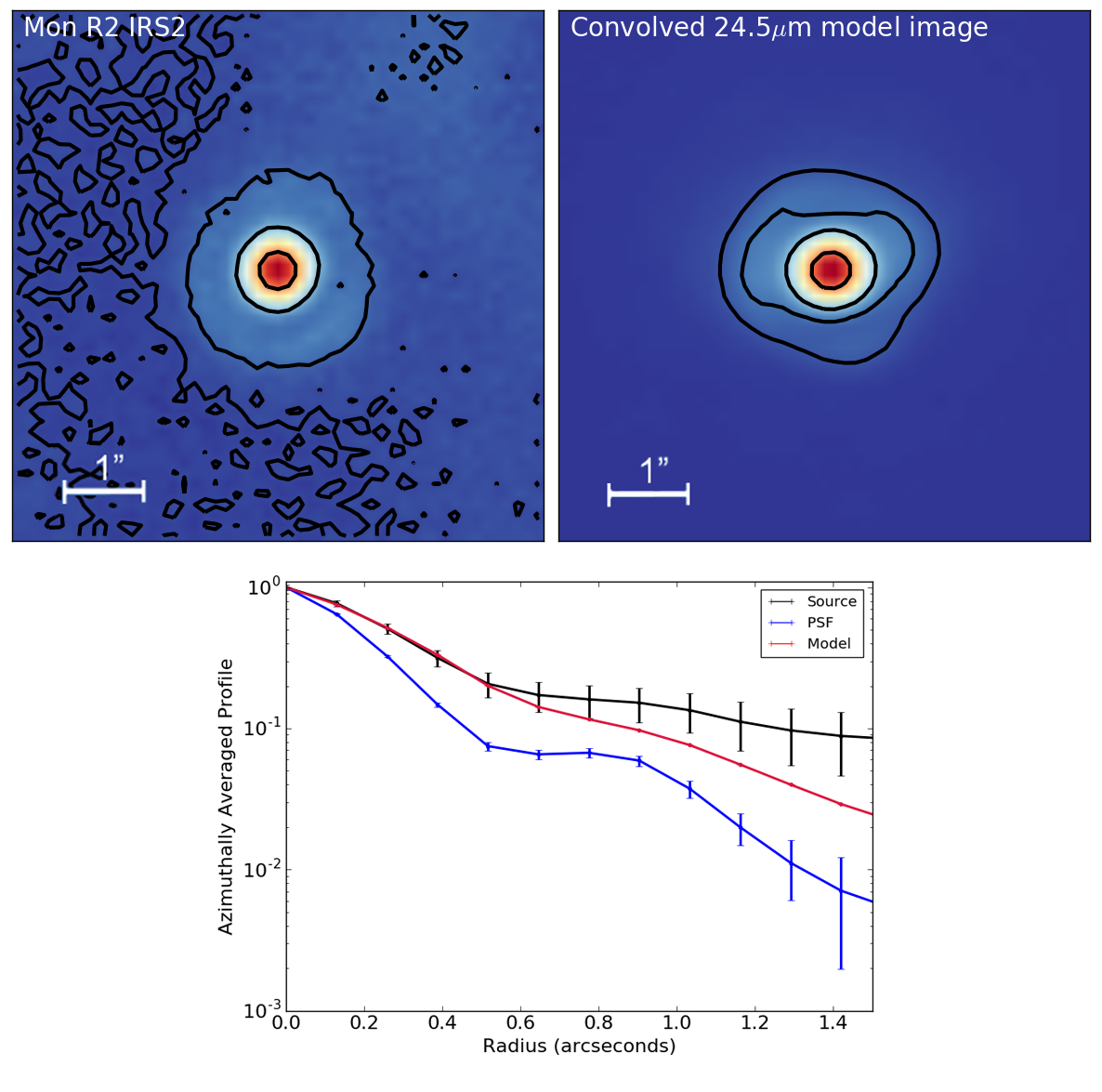}
   \caption{COMICS 24.5$\mu$m image (top left), convolved model image (top right) and subsequent radial profiles (bottom). The model image was convolved with the PSF of the observed object to accurately mimic the effects of the telescope specific to the observations. The contours in the images represent 5, 10, 25 and 75\% of the peak flux. The COMICS image is subject to a large amount of low level contamination, which can be seen in the 5\% contours. In order to avoid this emission, only the inner regions of the source were considered in the fit.}
   \label{monimgs}
   \end{figure*} 
         \begin{figure}[h!]
   \centering
   \includegraphics[width=90mm]{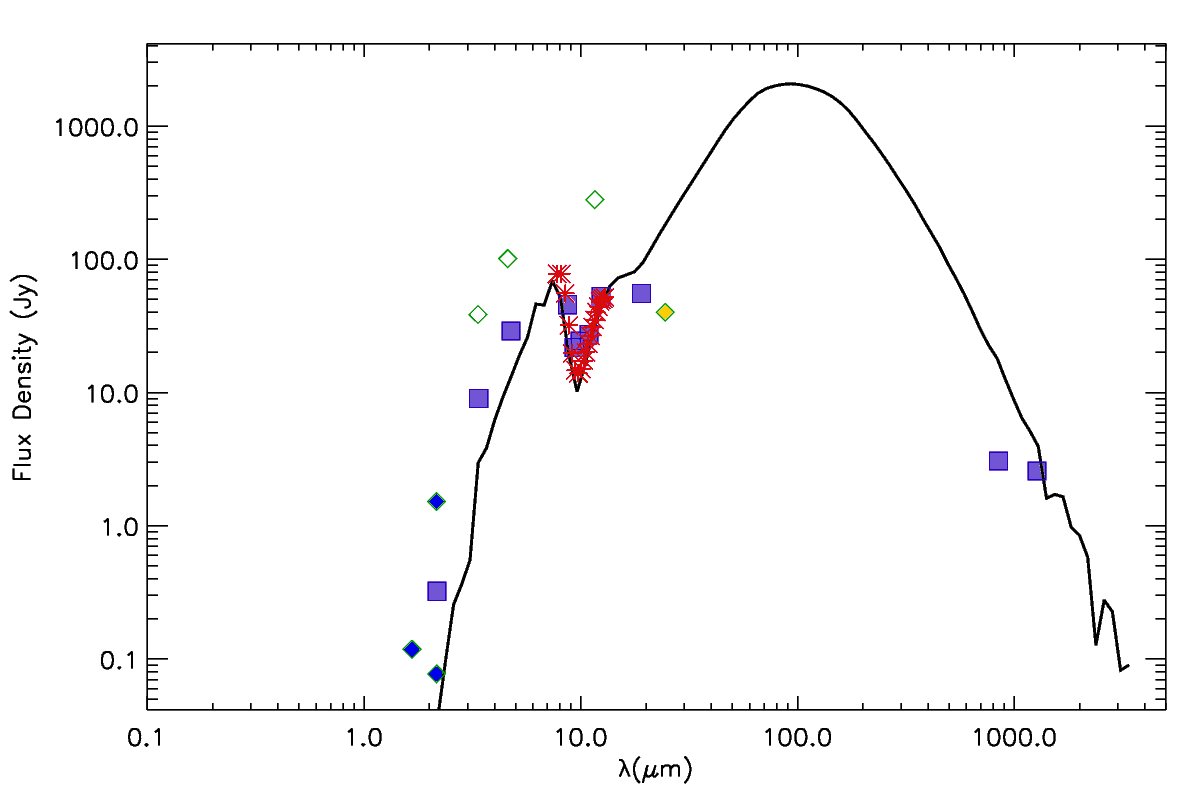}
   \caption{Model SED of the best-fitting model of Mon R2 IRS2 (black). Multi-wavelength flux measurements from the RMS are represented as blue diamonds, the yellow diamond represents the VISIR flux density and the fluxes corresponding to the MIDI visibilities are also shown in red. The unfilled diamonds represent the fluxes that were not considered in the fitting due to their suspected contamination and the fluxes from \citet{hen92} are included as squares.}
   \label{monsed}
   \end{figure} 

\subsection{Mon R2 IRS2}

The Mon R2 region is home to five different sources and located at a distance of $\sim$840pc \citep{mondist}. A large extended HII region, illuminated mostly by scattered light, covers 30" of the region and is visible in both 870$\mu$m maps \citep{giannakopoulou} and the 24.5$\mu$m image of \citet{witcomics}. Mon R2s IRS2 lies within the centre-north region of this H\textsc{ii} region and is its main source of illumination \citep{aspwalt}. A large CO bipolar outflow is present in the wider environment \citep{giannakopoulou} but it is unknown which MYSO in the region is responsible for it. Another massive protostellar system, Mon R2 IRS3, which has been resolved into two sources A and B, is the dominant source of the whole region. No free-free emission is associated with IRS2 and it is unresolved in UKIRT observations \citep{alv04}. \citet{boley13} find its mid-IR emission to be compact ($\sim$40au), and attribute this to its lower luminosity than IRS3A and B. \citet{hen92} find a luminosity of the source of $\sim$6000L$_{\odot}$. Recent work by \citet{monalma} studies the gas of Mon R2 IRS2. They find that their ALMA images, studies of the H21$\alpha$ emission and 3D radiative transfer modelling show the presence of an ionised Keplerian disk and a neutral rotating disk. The disk is highly flared (as shown in Figure 4 of the cited paper) and deviation of the H21$\alpha$ emission centroids from the disk plane imply warping in the disk, which they suggest could be due to the presence of a secondary object. 

Our work finds that Mon R2 IRS2 is another source where the fits to the MIDI data are improved by the inclusion of an inner hole (20au in radius). The dust sublimation radius of the source is small ($\sim$6au) due to comparatively low source luminosity required to fit the other observables. The PAH emission lines for the source are weak \citep{pahs}, and it was found that the MIDI fit could be improved by removing the PAHs from the disk. Removing the PAHs in the envelope and cavity material had negligible effects on the MIDI fits, implying that the N-band emission of the source is disk-dominated. Removing PAHs from the entire environment greatly worsens the NIR slope of the SED fit. 
The final luminosity of 5520L$_{\odot}$ is similar to that of \citet{hen92}. The envelope of Mon R2 IRS2 is similar to that of the majority of the sample, with a comparable infall rate and the same outer radius. The cavity density is low ($\sim$10$^{-21}$gcm$^{-3}$) and the cavity opening angle is 30$^{\circ}$. The source is inclined at 130$^{\circ}$ so both cavity lobes are visible. Through the SED fitting it was difficult to simultaneously fit the silicate absorption feature and the 20$\mu$m flux. Eventually the fitting of the silicate absorption feature was prioritised as this is the dataset with higher spatial resolution. The 
visibilities in Figure \ref{monvis}, the images in Figure \ref{monimgs} and the SED in Figure \ref{monsed}.

\begin{figure*}[h!]
   \centering
   \includegraphics[width=135mm]{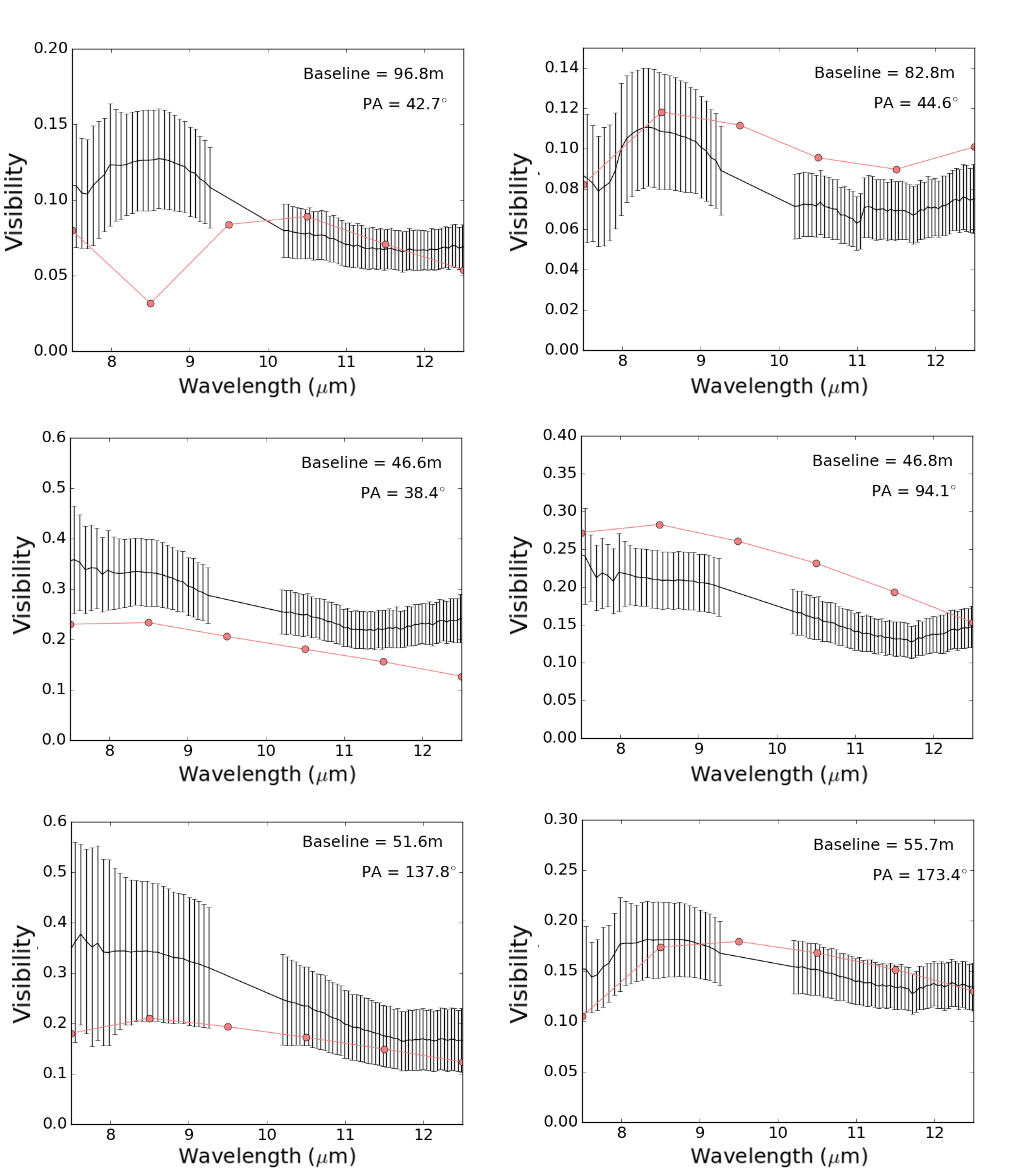}
   \caption{Observed visibilities for each configuration (black) with the simulated visibilities for each model image (coloured) for M8EIR.}
   \label{m8vis}
   \end{figure*} 
      \begin{figure*}
   \centering
   \includegraphics[width=140mm]{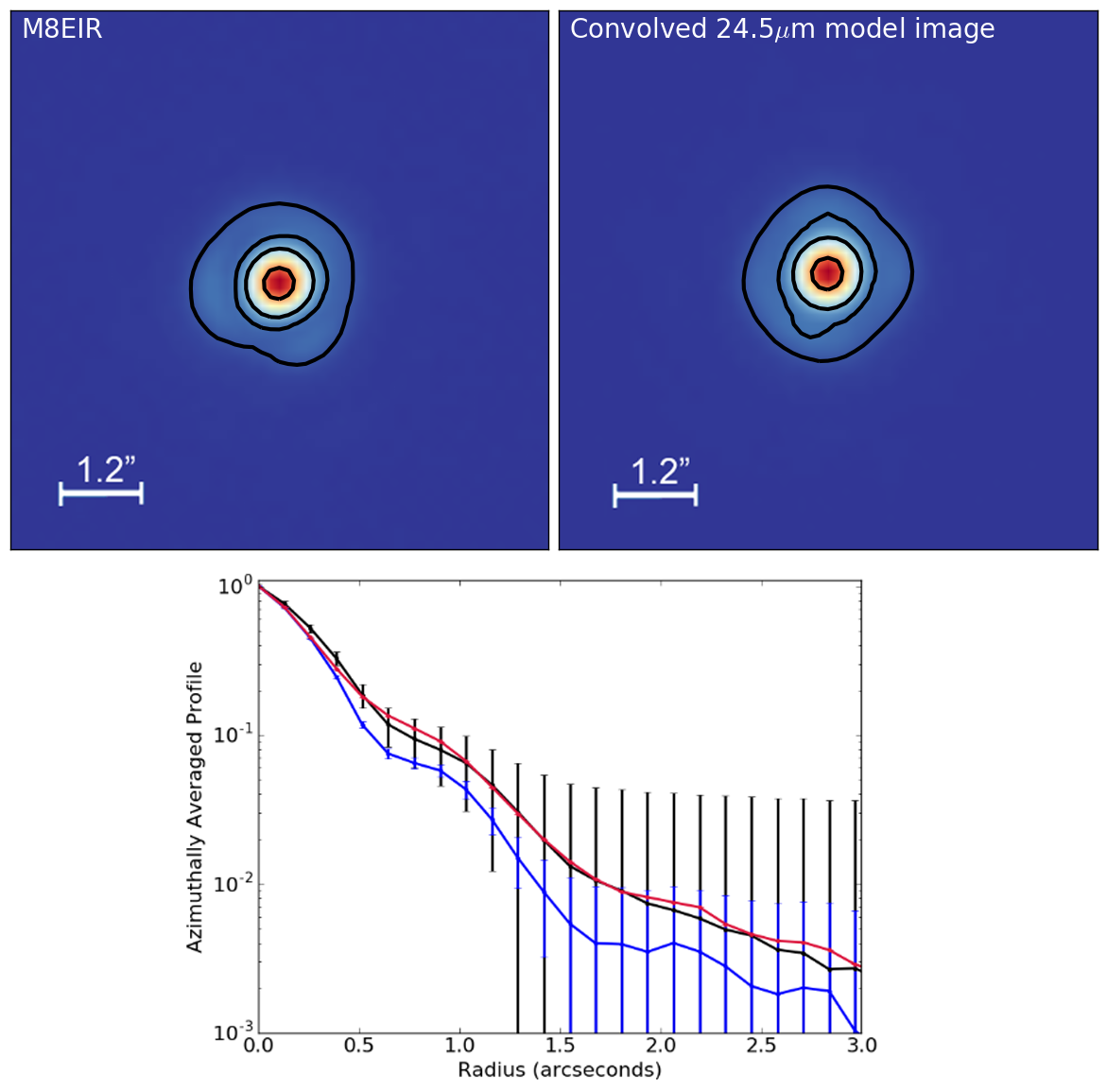}
   \caption{COMICS 24.5$\mu$m image (top left), convolved model image (top right) and subsequent radial profiles (bottom). The model image was convolved with the PSF of the observed object to accurately mimic the effects of the telescope specific to the observations. The contours in the images represent 5, 10, 25 and 75\% of the peak flux.}
   \label{m8imgs}
   \end{figure*} 
         \begin{figure}
   \centering
   \includegraphics[width=90mm]{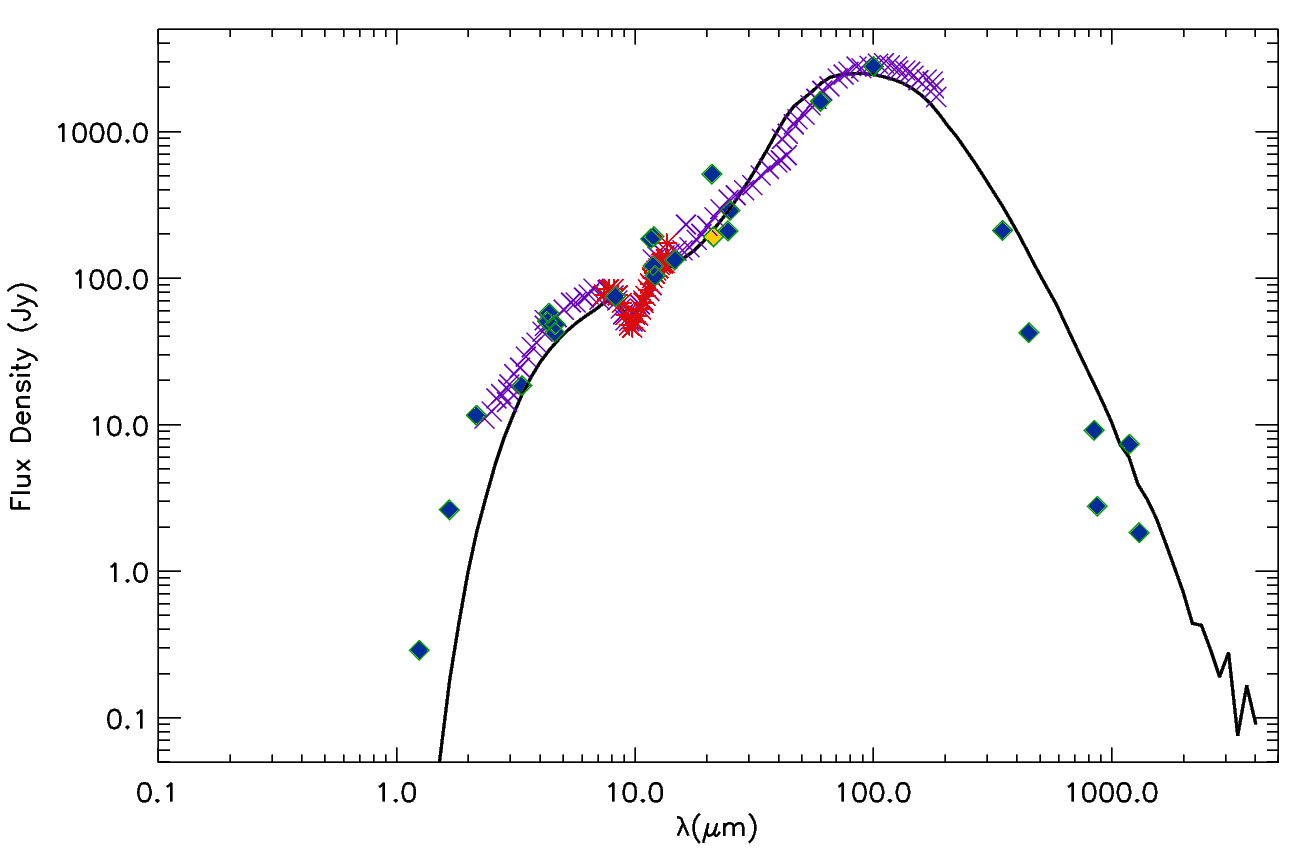}
   \hspace*{1cm}
   \caption{Model SED of the best-fitting model (black) for M8EIR. All fluxes from the literature are shown as diamonds, except for the spectrum from \citet{feldtm8}. This was sampled electronically and the data are represented as purple crosses. The fluxes corresponding to the MIDI visibilities are shown in red. Open symbols correspond to upper limits.}
   \label{m8sed}
   \end{figure} 
   
The position angles of the MIDI baselines for Mon R2 IRS2 mean that all five configurations are likely to be tracing disk material and configuration i) is likely to be tracing cavity material. Three of the MIDI configurations have high visibilities and are tracing an unresolved component of the protostellar environment. The simulated visibilities of configuration i) get lower if the outer radius of the disk is made larger implying that a small, compact disk best satisfies this configuration. 
However, as $R_c$/$R\SPSB{disk}{max}$ increased to values of 4000au (in agreement with some of the largest observed disks around MYSOs e.g. \citet{john}) improvements were seen to the fit of the COMICS profile as the flux at larger radii increased, so a compromise ultimately had to be made. 
Configurations iv) and v) of the MIDI data proved the most difficult to fit and resulted in the inclusion of the hole. Without this the visibilities for both the configurations were too high. This hole is 40au across, in agreement with the size of the compact emission from \citet{boley13}. This implies that the mid-IR emission for this source, as traced by their work, is mostly dominated by emission from this inner rim. A compromise was found between the fitting of configurations iv) and v) as opposed to fitting one perfectly and leaving the other poorly fit. 




\subsection{M8EIR}

This work finds that the 13.5M$_{\odot}$ central object of \citet{jilee13} well reproduces the shape of the SED of M8EIR. A luminosity of $\sim$12000L$_{\odot}$ is sufficient to fit the multiple datasets, which is also in good agreement with the value of $L$ obtained for the source in \citet{jilee13}. In the preferred model a 5$\times$10$^{5}$au envelope, infalling at a rate of 1$\times$10$^{-3}$M$_{\odot}$yr$^{-1}$, surrounds this protostar. The presence of an envelope agrees with previous work by \citet{simon84} who interpreted M8EIR as two physically different components,a hot compact one and a broader cooler one, and stated that the broad component is not molecular cloud material, but an `intermediary phase' of material between molecular cloud and hot disk material. 
A disk is present with a maximum radius of 2000au and outflow cavities are carved out of the envelope with opening angles of 25$^{\circ}$ and densities of order 10$^{-20}$gcm$^{-3}$. Using a cavity density exponent of 0.25 was found to improve both the simulated 24.5$\mu$m profile and the shape of the SED, similar to NGC 2264 IRS1 and IRAS 17216-3801. Using an inner hole (radius 30au) again improved the fit of the simulated visibilities. Using the dust sublimation radius ($\sim$15au) as the minimum radius led to the simulated visibilities for all configurations to be lower than those observed. In configuration i) a sharp downturn is found at 8.5$\mu$m. Such a quick visibility change can be caused by the turnover induced by the inner rim of the disk. When varying the inner disk radius around 30au similar shapes were observed, but alongside further detriments to the other visibility datasets. As a result 30au was chosen as the final inner radius. The simulated visibilities between 9-12$\mu$m for configuration ii) are $\sim$0.02 higher than those observed, implying that the structure traced by this configuration is slightly larger than that modelled. Given the long baseline length of the configuration this could be explained if the inner rim is not completely smooth but clumpy in nature. If this inner radius were the dust destruction radius a sharp cut off would be expected, but given that an inner hole exists the idea that the medium could also not be perfectly smooth is not unlikely. A compromise was found between the fitting of configurations iii) and iv) as opposed to fitting one perfectly and the other very poorly. 
The visibilities are shown in Figure \ref{m8vis}, the images in Figure \ref{m8vis} and the SED in Figure \ref{m8sed}. 

\begin{figure*}
   \centering
   \includegraphics[width=130mm]{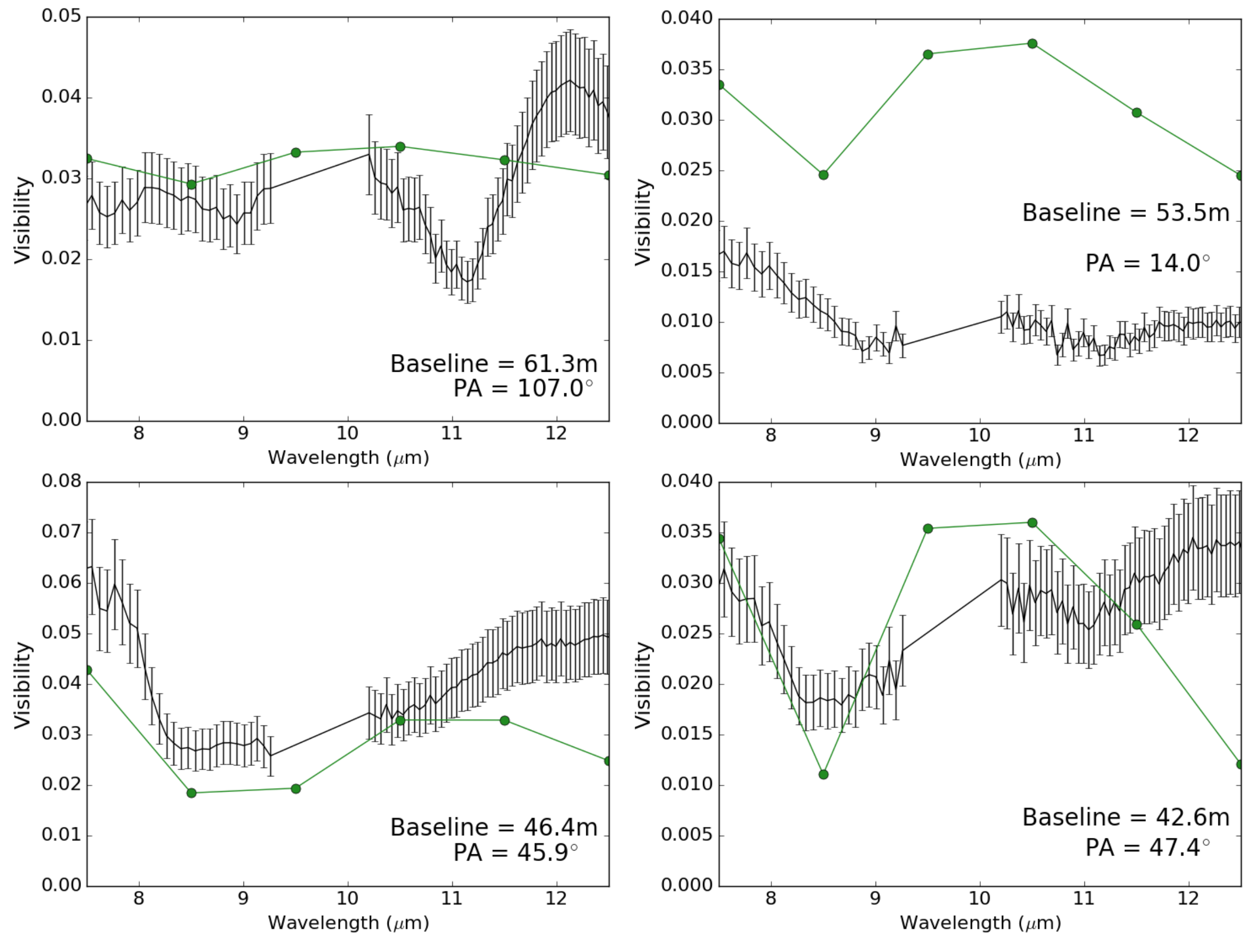}
   \caption{Observed visibilities for each configuration (black) with the simulated visibilities for each model image (coloured) for AFGL 2136.}
   \label{afglvis}
   \end{figure*} 
 
      \begin{figure*}
   \centering
   \includegraphics[width=130mm]{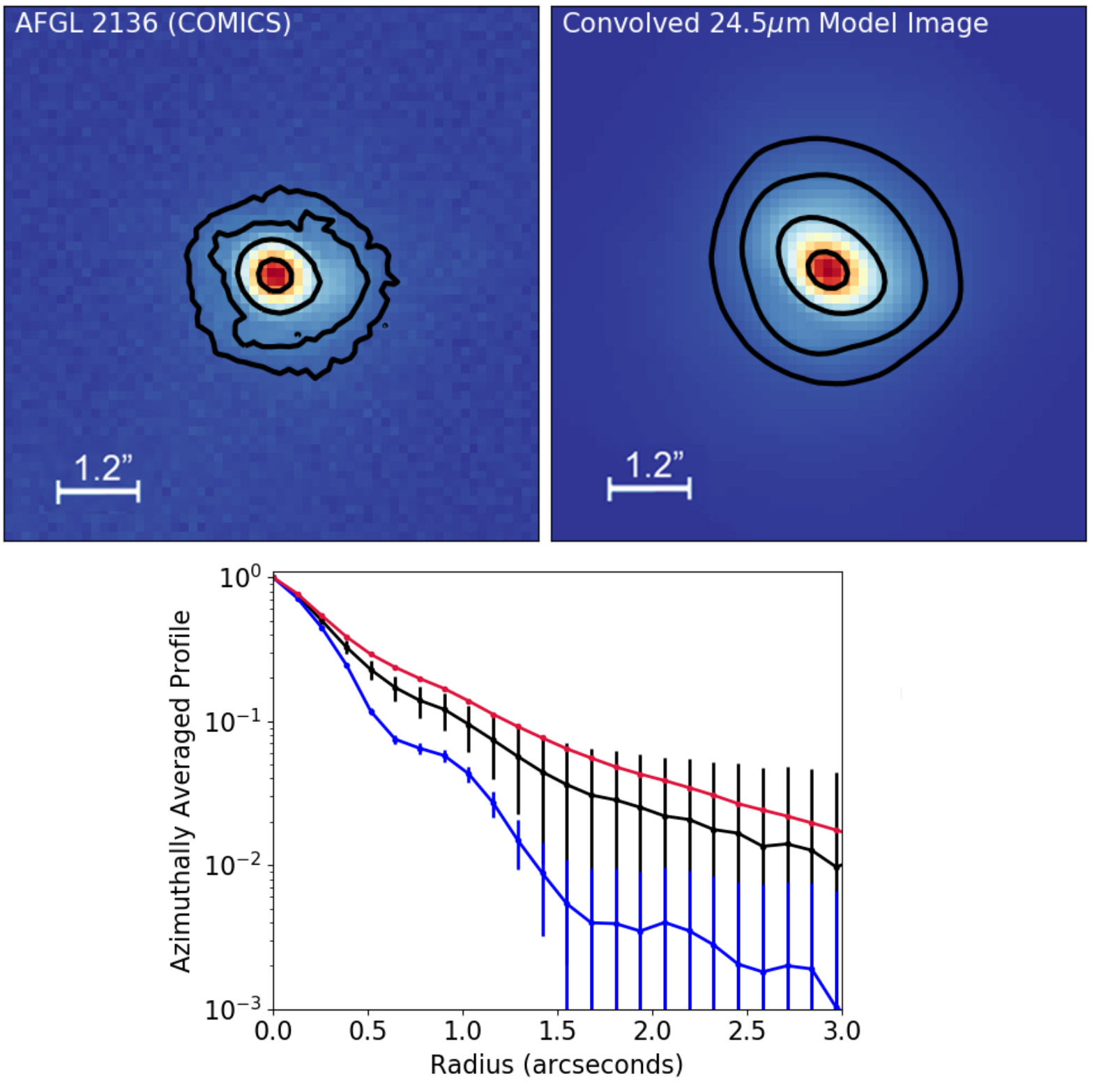}
   \caption{COMICS 24.5$\mu$m image (top left), convolved model image (top right) and subsequent radial profiles (bottom). The model image was convolved with the PSF of the observed object to accurately mimic the effects of the telescope specific to the observations. The model is more symmetric than the observed source meaning it has more emission in its eastern regions, explaining the larger amount of 24.5$\mu$m flux visible for the model in the radial profiles. The contours in the images represent 5, 10, 25 and 75\% of the peak flux.}
   \label{afglimgs}
   \end{figure*} 
   
         \begin{figure}
   \centering
   \includegraphics[width=90mm]{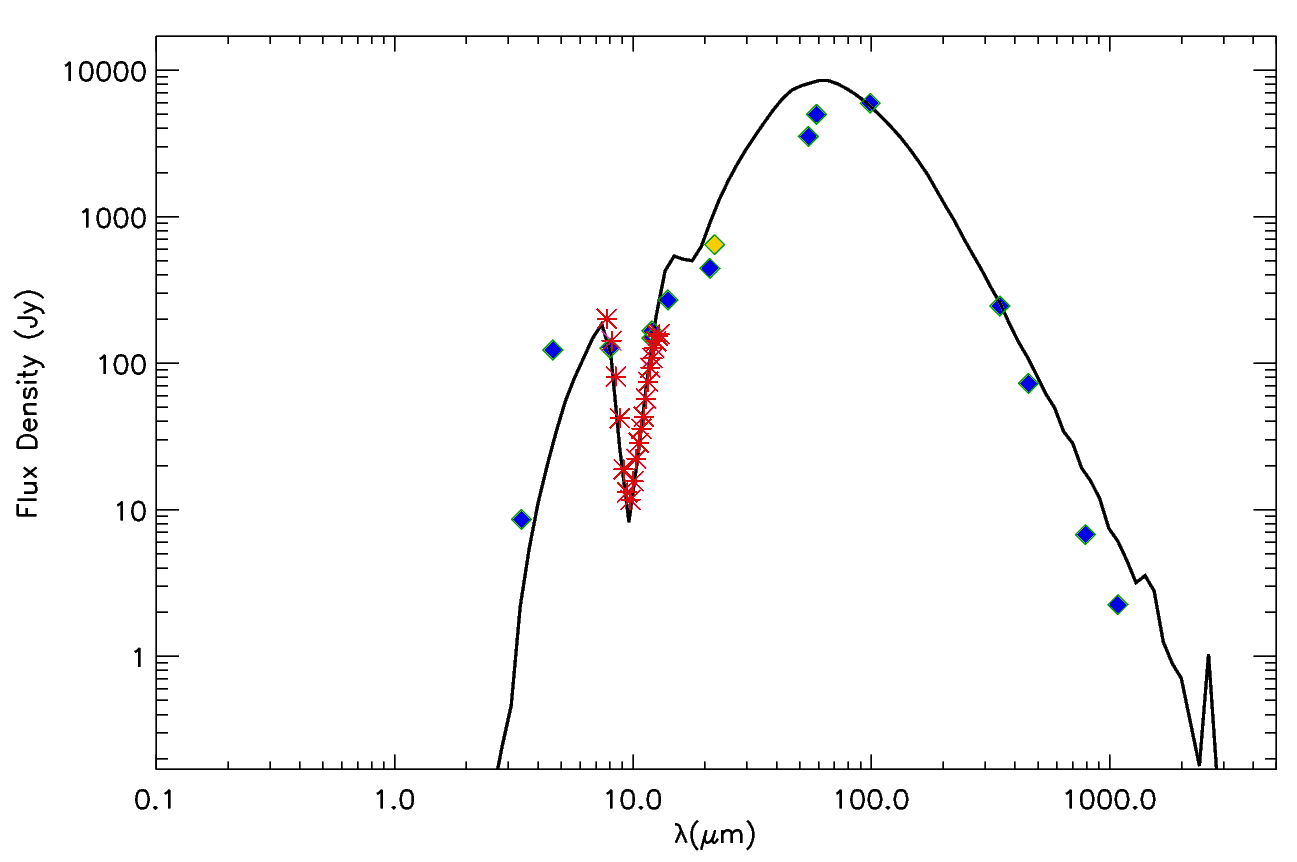}
   \caption{Model SED of the best-fitting model (black). Multi-wavelength flux measurements from the RMS are represented as blue diamonds, the yellow diamond represents the COMICS flux density and the fluxes corresponding to the MIDI visibilities are also shown in red. The unfilled diamonds represent the fluxes that were not considered in the fitting due to their suspected contamination.}
   \label{afglsed}
   \end{figure} 

\subsection{AFGL 2136}


AFGL 2136 IRS1 (hereafter AFGL 2136) was studied by \citet{wit11} using a similar analysis to our own. The main difference was the amount of MIDI data available as only one configuration was fit in that work, as opposed to the four used here. Additional technical differences between our fitting and \citet{wit11} are the multiplication of the model images with a Gaussian and the convolution of the COMICS image with a Gaussian PSF as opposed to an observed one (see Paper I). The COMICS image was presented in \citet{wit11} but not included in the fitting process. The final model of \citet{wit11} was used as a starting point for the modelling in our work.It was found that without multiplying by the Gaussian, a good fit could not be found to the MIDI configuration common to both this paper and \citet{wit11}. The model also provided poor fits to all the other MIDI configurations that were not included in \citet{wit11}.

\citet{maudafgl1} find that the dust emission implies a disk-like structure, so various disk-outflow-envelope geometries were trialled during the modelling. disk masses ranging between 10-30 solar masses are workable components of the model, in agreement with the kinematic measurements of \citet{maudafgl1} and \citet{maudafgl2}. The larger the mass, the less the 24.5$\mu$m emission, improving the fit of the radial profile. For a 30 solar mass disk the millimetre slope of the SED fit is worse than for a 10 solar mass disk. The quality of the MIDI fits are comparable across the disk mass range. For the 30 solar disk mass model, the fit of configuration ii) improves slightly as the simulated visibility at 8$\mu$m increases, but the visibilities for configuration iv) also increase, which further worsens the fit to that configuration by a comparable amount. Given the similarity across the MIDI data a compromise was made between the SED and COMICS fits and 20 solar masses was selected as the final disk mass. 

Similarity is found between the preferred best fitting model of our work and the model of \citet{wit11} in that the minimum dust radius is larger than that of the dust sublimation radius. 
In this work the inner radius is four times larger than the sublimation radius, in \citet{wit11} it was five. In \citet{wit11} a gaseous disk and bloated star resided within this cleared inner hole. Including a gaseous disk is not part of the methodology of this work, but it could exist within the inner dust rim. As discussed in the examination of G305.20+0.21 in Paper I, the presence of a bloated source can also not be ruled out as our observables are sensitive to overall luminosity of the central source, not its specific radius or temperature. The final model includes an inclination for the system of 60$^{\circ}$, and the central object is similar to that of \citet{wit11}; a bloated star of radius 25R$_{\odot}$ with an overall luminosity of 1.51$\times$10$^{5}$L$_{\odot}$. 
This luminosity matches the O-type classification ascribed by \citet{maudafgl1} and \citet{maudafgl2} and in good agreement with the bolometric luminosity determined by \citet{rmslum}. Both a large and small grains disk were used in the final model, with matching scale-heights. The outflow cavity opening angle is tightly constrained at 22.5$^{\circ}$ to maintain the balance of fits of the three observables and the cavity density is of order 10$^{-19}$gcm$^{-3}$. The centrifugal radius is 2000au and the infall rate is of order $10^{-4}$ solar masses per year. 

The visibilities of the final model are shown in Figure \ref{afglvis}, the images are shown in Figure \ref{afglvis} and the SED is shown Figure \ref{afglsed}. Configuration ii) of the MIDI data proved nigh-impossible to well fit. Despite this, the preferred model recreated the shape of the visibilities well and was only $V^2$=0.015 off in visibility. This implies that the protostellar environment traced by this configuration is larger than the model environment. The PA of this configuration was 14$^{\circ}$ meaning that, taking into account the source's PA on the sky, it could be tracing envelope material close to the cavity wall or disk material. Both the continuum and CO contours of \citet{maudafgl1} show the medium to be clumpy, despite the disk's stability. Given that only smooth geometries were used in the modelling, it could be that the model is not reproducing this emission leading to this poor fitting. The other three MIDI configurations are matched well by the final geometry. 

A further study at longer wavelengths, \citet{maudafgl2}, detect a ring-like structure at scales of 120au. This is in very good agreement with the inner dust radius found in our model, which is four times the sublimation radius or $\sim$125au. 

\section{List of MIDI observations}

\begin{table*}
\caption{List of MIDI observations of for the sample. The starred dataset is data from \citet{grell}. The data for W33A are from \citet{wit10}. All other data are from \citet{boley13}.}               
\label{midiobs}      
\centering                                      
\begin{tabular}{c c c c c c c}          
\hline\hline                        
Object & Configuration & Date & Telescopes & Projected baseline &  Position angle & ESO Run ID  \\    
& & (UTC) & & (m) & ($^{\circ{}}$) & \\
\hline                                   
G305.20+0.21 & i) & 2005-06-26 &  UT1-UT2  & 42.5  & 48.1 & 75.C-0755(B) \\
& ii) & 2005-02-27  & UT2-UT3  & 44.7  & 4.8 &  74.C-0389(A) \\
& iii) & 2005-03-02  & UT3-UT4 &  56.7  & 83.4  & 74.C-0389(B) \\
& iv) & 2005-03-02  & UT3-UT4  & 58.9 &  93.4  &  74.C-0389(B) \\
& v) & 2005-03-02  & UT3-UT4  & 61.0  & 106.9  &  74.C-0389(B) \\
& vi) & 2005-06-24 &  UT3-UT4 &  62.3  & 146.7  &  75.C-0755(A) \\
\hline                           
W33A & i) & 2008-06-21 &  UT2-UT3  & 42.03  & 15.04  & 381.C-0602 \\
 & ii) & 2008-04-22  & UT3-UT4  & 57.95  & 119.69 & 381.C-0602 \\
 & iii) & 2008-04-21  & UT3-UT4 &  61.08  & 114.57 & 381.C-0602\\
 & iv) & 2005-09-16  & UT2-UT3  & 45.58 &  47.31 & 381.C-0602\\
\hline  
NGC 2264 IRS1 & i) & 02/15/2009 & UT2-UT3 & 40.2 & 43.9 & 082.C-899(A) \\
 & ii) & 12/23/2005 & UT2-UT4 & 89.1 & 81.1 & 076.C-0725(B) \\
\hline
S255 IRS3 & i) & 2010-01-15 & E0-G0 & 15.8 & 74.8 & 84.C-0183(D) \\
 & ii) & 2010-01-16 & E0-H0 & 47.8 & 73.5 & 84.C-0183(C) \\
 & iii) & 2010-01-17 & G0-H0 & 30.8 & 76.7 & 84.C-0183(B) \\
 & iv) & 2010-01-17 & G0-H0 & 31.1 & 76.0 & 84.C-0183(B) \\
 & v) & 2010-01-17 & E0-H0 & 47.5 & 68.9 & 84.C-0183(B) \\
 & vi) & 2013-02-13 & C1-D0 & 17.6 & 32.6 & 90.C-0717(A) \\
\hline
IRAS 17216-3801 & i) & 2012-05-01 & B2-C1 & 11.3 & 13.9 & 89.C-0968(A) \\
 & ii) & 2012-05-01 & A1-B2 & 11.0 & 134.1 & 89.C-0968(A) \\
 & iii) & 2012-05-01 & B2-C1 & 10.0 & 40.2 & 89.C-0968(A) \\
 & iv) & 2012-05-02 & A1-B2 & 8.9 & 91.5 & 89.C-0968(A) \\
 & v) & 2012-05-02 & A1-C1 & 16.0 & 69.6 & 89.C-0968(A) \\
 & vi) & 2012-05-02 & A1-D0 & 34.2 & 51.5 & 89.C-0968(A) \\ 
 & vii) & 2012-05-02 & B2-D0 & 30.5 & 39.5 & 89.C-0968(A) \\
 & viii) & 2013-02-12 & B2-D0 & 33.8 & 165.1 & 90.C-0717(A) \\
 & ix) & 2013-02-12 & C1-D0 & 22.6 & 175.5 & 90.C-0717(A) \\
 & x) & 2013-02-13 & A1-D0 & 35.1 & 176.9 & 90.C-0717(A) \\
 \hline
Mon R2 IRS2 & i) & 2009-02-15 & UT2-UT3 & 44.6 & 39.9 & 82.C-0899(A) \\
 & ii) & 2009-11-14 & E0-G0 & 10.8 & 55.8 & 84.C-1072(A) \\
 & iii) & 2009-11-15 & E0-G0 & 10.7 & 55.4 & 84.C-1072(A) \\
 & iv) & 2009-11-15 & G0-H0 & 26.9 & 64.9 & 84.C-1072(B) \\
 & v) & 2009-11-15 & G0-H0 & 31.1 & 70.6 & 84.C-1072(B) \\
\hline
M8EIR & i) & 2004-06-05 & UT1-UT3 & 96.8 & 42.7 &  60.A-9224(A) \\
 & ii) & 2004-06-05 & UT1-UT3 & 82.8 & 44.6 & 60.A-9224(A) \\
 & iii) & 2004-08-01 & UT2-UT3 & 46.6 & 38.4 & 273.C-5044(A) \\
 & iv) & 2005-03-02 & UT3-UT4 & 46.8 & 94.1 & 74.C-0389(B) \\
 & v) & 2005-06-24 & UT3-UT4 & 51.6 & 137.8 & 75.C-0755(A) \\
 & vi) & 2005-06-26 & UT1-UT2 & 55.7 & 173.4 & 75.C-0755(B) \\
\hline
AFGL 2136 & i) & 2005-06-24 & UT3-UT4 & 61.3 & 107.0 &  75.C-0755(A) \\
 & ii) & 2005-06-26 & UT1-UT2 & 53.5 & 14.0 & 75.C-0755(B) \\
 & iii) & 2006-05-18 & UT2-UT3 & 46.4 & 45.9 & 77.C-0440(A) \\
 & iv) & 2008-06-23 & UT2-UT3 & 42.6 & 47.4 & 381.C-0607(A) \\
\hline
\end{tabular}
\end{table*}

\section{List of SED fluxes}

\begin{table*}
\caption{Fluxes available for use in the SED fitting of W33A.}            
\label{w33asedtab}      
\centering      
\begin{tabular}{c c c}           
\hline\hline                         
Origin & Wavelength ($\mu$m) & Flux (Jy) \\
\hline          
2MASS & 1.235 & (1.21$\pm$0.07)$\times$10$^{-3}$\\
 & 1.66 & (5.37$\pm$0.3)$\times$10$^{-3}$\\
 & 2.159 & 0.139$\pm$0.01\\
WISE & 3.4 & 0.714$\pm$0.001 \\
& 4.6 & 20.8$\pm$3 \\
& 12 & 33.3$\pm$2 \\
MSX & 8 & 15.6$\pm$4 \\
& 12 & 23.2$\pm$5 \\
& 14 & 50.0$\pm$6 \\
& 21 & 144$\pm$6 \\
IRAS & 60 & 2230$\pm$500 \\
MIPS & 70 & 2137$\pm$6 \\
SHARC & 350 & 597$\pm$30 \\
JCMT & 850 & 40.5$\pm$0.11 \\
SIMBA & 1200 & 15.7$\pm$3 \\
\hline                                             
\end{tabular}
\end{table*}

\begin{table*}
\caption{Fluxes available for use in the SED fitting of NGC 2264 IRS1.}            
\label{sedtabngc}      
\centering      
\begin{tabular}{c c c c}           
\hline\hline                         
Origin & Wavelength ($\mu$m) & Flux (Jy) & Reference \\
\hline                                    
2MASS & 1.24 & 0.040$\pm$0.002 & \citet{2mass} \\
& 1.66 & 0.898$\pm$0.03 & \\
& 2.16 & 7.15$\pm$0.04 &\\
\hline    
WISE & 3.35 & 12.1$\pm$1 & \citet{wise} \\
& 4.6 & 132 & \\
& 11.6 & 127$\pm$40&\\
& 22 & 353$\pm$0.4 &\\
\hline    
MSX & 8 & 91.4$\pm$4 & \citet{egan} \\
& 12 & 119$\pm$6 &\\
& 14 & 148$\pm$9 &\\
& 21 & 13.7$\pm$14 &\\
\hline    
COMICS & 24.5 & 330$\pm$33 & \citet{witcomics} \\
\hline
KAO & 53 & 980$\pm$5 & \citet{harv} \\
KAO & 100 & 1645$\pm$82 &\\
KAO & 175 & 1530$\pm$77 &\\
\hline
IRTF & 350 & 188$\pm$7 & \citet{chini}\\
\hline                                             
\end{tabular}
\end{table*}

\begin{table*}
\caption{Fluxes available for use in the SED fitting of S255 IRS3.}            
\label{sedtabs255}      
\centering      
\begin{tabular}{c c c c}           
\hline\hline                         
Origin & Wavelength ($\mu$m) & Flux (Jy) & Reference\\
\hline 
NICMOS & 2 & 0.00283$\pm$0.0003 & \citet{simp} \\
\hline 
CIAO (Subaru) & 2.2 & 0.010$\pm$0.0001 & \citet{itoh} \\
 & 3.75 & 2.38$\pm$0.002 & \\
 & 4.70 & 13.1$\pm$0.013 &  \\
 \hline 
Michelle (Gemini) & 7.9 & 79.5$\pm$0.7 & \citet{long06} \\ 
& 8.8 & 21.8$\pm$0.2 &  \\ 
& 11.6 & 47.7$\pm$0.4 &  \\ 
& 12.5 & 95.3$\pm$1 & \\ 
\hline 
UKIDSS & 1.64 & 0.00865$\pm$0.0005 & \citet{ukidss} \\
 & 2.16 & 0.032$\pm$0.001 & \\
\hline 
ISACC & 1.66 & 0.00237$\pm$0.004 & \citet{car18} \\
 & 2.16 & 0.0153$\pm$0.0002 &  \\
\hline 
IRAC & 3.6 & 2.62$\pm$0.01 & \citet{fazio} \\
 & 4.5 & 10.91$\pm$0.03 & \\
 & 5.8 & 34.5$\pm$0.04 & \\
 & 8.0 & 41.2$\pm$0.04 & \\
\hline 
AKARI/IRC & 8.6 & 103$\pm$0.3 & \citet{jaxa} \\
 & 18.4 & 213$\pm$0.2 & \\
 \hline 
AKARI & 65 & 455$\pm$10 & \citet{car18} \\
 & 160 & 760$\pm$3 & \\
\hline
PACS & 70 & 1465$\pm$150 & \citet{pacs} \\
& 160 & 849$\pm$80 & \\
\hline
SHARC & 350 & 118$\pm$9 & \citet{sharc} \\
\hline
IRAM & 850 & 0.5$\pm$0.07 & \citet{s255zin} \\
SMA & 1050 & 0.45$\pm$0.06 & \citet{zin12} \\
SMA & 1300 & 0.058$\pm$0.01 & \citet{s255zin} \\
\hline
2MASS & 1.24 & (1.01$\pm$0.2)$\times$10$^{-3}$ & \citet{2mass} \\
& 1.66 & (6.46$\pm$1)$\times$10$^{-3}$ & \\
 & 2.16 & (5.06$\pm$0.5)$\times$10$^{-2}$ &\\
\hline 
WISE & 3.35 & 0.689$\pm$0.01 & \citet{wise} \\  
& 4.6 & 11.7$\pm$0.1 \\
& 11.6 & 222$\pm$70\\
& 22 & 695$\pm$0.8\\
\hline 
MSX & 8 & 55.0$\pm$2 & \citet{egan} \\
& 12 & 104.0$\pm$5 &\\
& 14 & 154$\pm$9 &\\
& 21 & 213$\pm$10 &\\
\hline 
COMICS & 24.5 & 163$\pm$20 & \citet{witcomics}\\
\hline 
UKIRT & 350 & 230$\pm$9 & \citet{richardson} \\
& 370 & 180$\pm$2 & \\ 
& 760 & 36$\pm$2 &  \\
& 1070 & 17$\pm$1 & \\
\hline
SCUBA & 450 & 224$\pm$0.2 & \citet{scuba} \\
& 850 & 22.0$\pm$0.02 & \\
\hline
IRTF & 350 & 190 & \citet{mez} \\
 & 1300 & 3.3$\pm$0.8 & \citet{chini2} \\
 & 1300 & 7.5 & \citet{mez} \\
\end{tabular}
\end{table*}

\begin{table*}
\caption{Fluxes available for use in the SED fitting of IRAS 17216-3801.}            
\label{sedtabafgliras}      
\centering      
\begin{tabular}{c c c c}           
\hline\hline                         
Origin & Wavelength ($\mu$m) & Flux (Jy) & Reference \\
\hline           
2MASS J-Band & 1.24 & 0.0134$\pm$0.0005 & \citet{2mass}\\
H-Band & 1.66 & 0.666$\pm$0.04 \\
Ks-Band & 2.16 & 6.73$\pm$0.1 \\
\hline  
WISE & 3.35 & 14.9$\pm$0.9 & \citet{wise}\\
& 4.60 & 52.5$\pm$20 \\
& 11.6 & 108$\pm$20 \\
& 22.1 & 430$\pm$4 \\
\hline  
MSX & 4.29 & 29.6$\pm$3 & \citet{egan}\\
& 4.35 & 50.1$\pm$5 \\
& 8.28 & 78.1$\pm$3 \\
& 12.1 & 145$\pm$7 \\
& 14.7 & 190$\pm$10 \\
& 21.34 & 299$\pm$20 \\
\hline  
VISIR & 19.5 & 245$\pm$10 & This work\\
\hline  
ATLASGAL & 870 & 35.1$\pm$5 & \citet{urq14} \\
\hline  
IRAS & 12 & 148$\pm$7 & \citet{iras} \\
& 25 & 410$\pm$20 \\
& 60 & 1640$\pm$200 \\
& 100 & 2820$\pm$400 \\
\hline                                             
\end{tabular}
\end{table*}

\begin{table*}
\caption{Fluxes available for use in the SED fitting of Mon R2 IRS2.}            
\label{sedtabmon}      
\centering      
\begin{tabular}{c c c c}           
\hline\hline                         
Origin & Wavelength ($\mu$m) & Flux (Jy) & Reference \\
\hline              
UKIDSS H-band & 1.662 & 0.002 (2$\pm$0.006)$\times$10$^{-3}$ & \citet{ukidss}\\
K-band & 2.159 & 0.0771$\pm$0\\
\hline  
WISE & 3.35 & 38.5$\pm$5 & \citet{wise} \\
& 4.60 & 101$\pm$15 \\
& 11.6 & 281$\pm$30 \\
& 22.1 & 4935.874 (upper limit) \\
\hline  
COMICS & 24.5 & 40$\pm$4 & \citet{witcomics}\\
\hline  
MRT & 870 & 3070$\pm$122 & \citet{hen92} \\
& 1300 & 2640$\pm$37 & \\
\hline                                                      
\end{tabular}
\end{table*}

\begin{table*}
\caption{Fluxes available for use in the SED fitting of M8EIR (1).}            
\label{sedtabm81}      
\centering      
\begin{tabular}{c c c c}           
\hline\hline                         
Origin & Wavelength ($\mu$m) & Flux (Jy) & Reference \\
\hline   
2MASS J-Band & 1.24 & (3.05$\pm$0.008)$\times$10$^{-3}$ & \citet{2mass}\\
H-Band & 1.66 & 2.62$\pm$0.1 \\
Ks-Band & 2.16 & 11.6$\pm$0.4 (poor quality (E flag)) \\
\hline   
MSX & 4.29 & 50.7$\pm$5 & \citet{egan}\\
& 4.35 & 57.6$\pm$6 \\
& 8.28 & 75.0$\pm$7 \\
& 12.1 & 104$\pm$10 \\
& 14.7 & 133$\pm$10 \\
& 21.34 & 191$\pm$10 \\
\hline   
COMICS & 24.5 & 210$\pm$20 & \citet{witcomics} \\
\hline   
IRTF & 1.24 & (6.7$\pm$0.07)$\times$10$^{-2}$ & \citet{simon84} \\
& 1.66 & 0.83$\pm$0.008 & \\
& 2.16 & 5.7$\pm$0.06 & \\
\hline  
SHARC &12 & 119$\pm$7 & \citet{mueller} \\
IRAS & 25 & 289$\pm$20 & \\
& 60 & 1611$\pm$200 & \\
& 100 & 2783$\pm$700 & \\
FCRAO & 350 & 210$\pm$42 & \citet{kast} \\
& 450 & 42.4$\pm$15 &\\
& 850 & 9.14$\pm$1 & \\
\hline  
SEST & 1200 & 7.36 & \citet{belt06} \\
\hline  
MRT & 870 & 2.79$\pm$0.1 & \citet{gurt} \\
SEST & 1300 & 1.85$\pm$0.02 & \\
\hline                                                      
\end{tabular}
\end{table*}

\begin{table*}
\caption{Fluxes available for use in the SED fitting of M8EIR (2).}            
\label{sedtabm82}      
\centering      
\begin{tabular}{c c c c}           
\hline\hline                         
Origin & Wavelength ($\mu$m) & Flux (Jy) & Reference \\
\hline   
ISO SWS & 16.4 & 232$\pm$70 & \citet{feldtm8} \\
ISO LWS & 2.34 & 11.0$\pm$2 & \\
& 2.50 & 12.3$\pm$2 &\\
& 2.62 & 15.1$\pm$3 &\\
& 2.74 & 16.2$\pm$3 &\\
& 2.90 & 14.3$\pm$3 &\\
& 2.94 & 17.3$\pm$3 &\\
& 2.97 & 18.8$\pm$4 &\\
& 3.11 & 22.1$\pm$4 &\\
& 3.29 & 24.5$\pm$5 &\\
& 3.48 & 29.1$\pm$6 &\\
& 3.69 & 33.8$\pm$7 &\\
& 3.86 & 36.2$\pm$7 &\\
& 4.09 & 43.0$\pm$9 & \\
& 4.18 & 47.2$\pm$9 & \\
& 4.19 & 51.8$\pm$10 & \\
& 4.43 & 49.9$\pm$10 & \\
& 4.69 & 54.1$\pm$11 & \\
& 5.03 & 60.7$\pm$12 & \\
& 5.38 & 68.1$\pm$13 & \\
& 6.11 & 73.8$\pm$15 & \\
& 6.39 & 77.2$\pm$15 & \\
& 6.69 & 84.7$\pm$17 & \\
& 7.09 & 71.2$\pm$14 & \\
& 7.34 & 75.4$\pm$15 & \\
& 7.68 & 82.7$\pm$17 & \\
& 8.13 & 76.2$\pm$15 & \\
& 8.42 & 70.3$\pm$14 & \\
& 8.72 & 61.9$\pm$12 & \\
& 8.92 & 56.4$\pm$11 & \\
& 9.13 & 51.5$\pm$10 & \\
& 9.45 & 50.9$\pm$10 & \\
& 9.78 & 50.3$\pm$10 & \\
& 10.0 & 56.4$\pm$11 & \\
& 10.5 & 66.3$\pm$13 & \\
& 10.8 & 70.2$\pm$14 & \\
& 11.5 & 81.5$\pm$16 & \\
& 11.7 & 89.4$\pm$18 & \\
& 11.9 & 134$\pm$27 & \\
& 12.4 & 140$\pm$28 & \\
& 13.3 & 143$\pm$29 & \\
& 13.9 & 147$\pm$29 & \\
& 14.9 & 153$\pm$31 & \\
& 15.8 & 155$\pm$31 & \\
& 16.9 & 162$\pm$32 & \\
& 18.1 & 182$\pm$36 & \\
& 19.0 & 200$\pm$40 & \\
\hline                                                      
\end{tabular}
\end{table*}

\begin{table*}
\caption{Fluxes available for use in the SED fitting of M8EIR (3).}            
\label{sedtabm83}      
\centering      
\begin{tabular}{c c c c}           
\hline\hline                         
Origin & Wavelength ($\mu$m) & Flux (Jy) & Reference \\
\hline   
ISO LWS & 20.1 & 227$\pm$45 & \citet{feldtm8}\\
& 21.5 & 260$\pm$52 & \\
& 22.8 & 295$\pm$59 & \\
& 24.7 & 343$\pm$69 & \\
& 26.5 & 367$\pm$73 & \\
& 28.7 & 398$\pm$80 & \\
& 31.1 & 431$\pm$86 & \\
& 33.7 & 500$\pm$100 & \\
& 36.5 & 555$\pm$100 & \\
& 39.5 & 615$\pm$100 & \\
& 40.9 & 637$\pm$100 & \\
& 43.3 & 690$\pm$100 & \\
& 41.3 & 889$\pm$200 & \\
& 43.8 & 963$\pm$200 & \\
 & 46.4 & 1120$\pm$200 & \\
& 51.4 & 1300$\pm$200 & \\
& 57.6 & 1670$\pm$300 & \\
& 54.4 & 1510$\pm$300 & \\
& 48.0 & 1210$\pm$200 & \\
& 61.0 & 1900$\pm$400 & \\
& 65.4 & 2030$\pm$400 & \\
& 70.8 & 2300$\pm$500 & \\
& 75.0 & 2440$\pm$500 & \\
& 78.5 & 2560$\pm$500 & \\
& 85.1 & 2770$\pm$600 & \\
& 89.0 & 2700$\pm$500 & \\
& 96.5 & 2800$\pm$600 & \\
& 105 & 2900$\pm$600 & \\
& 111 & 2930$\pm$600 & \\
& 117 & 2830$\pm$600 & \\
& 125.6 & 2764$\pm$600 & \\
& 127 & 2610$\pm$500 & \\
& 136 &	2610$\pm$500 & \\
& 143 & 2550$\pm$500 & \\
& 147 & 2400$\pm$500 & \\
& 154 & 2240$\pm$400 & \\
& 167 & 22670$\pm$500 & \\
& 177 &	2190$\pm$400 & \\
& 179 & 1970$\pm$400 & \\
& 184 & 1720$\pm$300 & \\
\hline                                                      
\end{tabular}
\end{table*}

\begin{table*}
\caption{Fluxes available for use in the SED fitting of AFGL 2136.}            
\label{sedtabafgl}      
\centering      
\begin{tabular}{c c c c}           
\hline\hline                         
Origin & Wavelength ($\mu$m) & Flux (Jy) & Reference \\
\hline          
2MASS J-Band & 1.24 & 1.59$\times$10$^{-3}$ & \citet{2mass}\\
H-Band & 1.66 & (8.52$\pm$0.01)$\times$10$^{-2}$ & \\
Ks-Band & 2.16 & 0.123$\pm$0.001 &\\
\hline 
WISE & 3.35 & 8.6$\pm$1 & \citet{wise}\\
& 4.6 & 123 & \\
& 11.6 & 148$\pm$40 &\\
& 22 & 644$\pm$1 &\\
\hline 
MSX & 8 & 127$\pm$5 & \citet{egan}\\
& 12 & 166$\pm$8 & \\
& 14 & 270$\pm$20 & \\
& 21 & 444$\pm$30 & \\
\hline 
COMICS & 24.5 & 140$\pm$10 & \citet{witcomics}\\
\hline 
IRAS & 60 & 4810$\pm$900 & \citet{iras}\\
& 100 & 5700$\pm$1000 & \\
\hline 
SHARC & 350 & 240$\pm$50 & \citet{mueller} \\
& 450 & 72.7$\pm$6 & \\
& 800 & 6.66$\pm$0.15 & \citet{kast} \\
& 1100 & 2.35$\pm$0.09 & \\
\hline                                             
\end{tabular}
\end{table*}

\end{appendix}

\end{document}